# On Full Diversity Space-Time Block Codes with Partial Interference Cancellation Group Decoding

Xiaoyong Guo and Xiang-Gen Xia

**Abstract**

In this paper, we propose a partial interference cancellation (PIC) group decoding for linear dispersive space-time block codes (STBC) and a design criterion for the codes to achieve full diversity when the PIC group decoding is used at the receiver. A PIC group decoding decodes the symbols embedded in an STBC by dividing them into several groups and decoding each group separately after a linear PIC operation is implemented. It can be viewed as an intermediate decoding between the maximum likelihood (ML) receiver that decodes all the embedded symbols together, i.e., all the embedded symbols are in a single group, and the zero-forcing (ZF) receiver that decodes all the embedded symbols separately and independently, i.e., each group has and only has one embedded symbol, after the ZF operation is implemented. The PIC group decoding provides a framework to adjust the complexity-performance tradeoff by choosing the sizes of the information symbol groups. Our proposed design criterion (group independence) for the PIC group decoding to achieve full diversity is an intermediate condition between the loosest ML full rank criterion of codewords and the strongest ZF linear independence condition of the column vectors in the equivalent channel matrix. We also propose asymptotic optimal (AO) group decoding algorithm which is an intermediate decoding between the MMSE decoding algorithm and the ML decoding algorithm. The design criterion for the PIC group decoding can be applied to the AO group decoding algorithm too. It is well-known that the symbol rate for a full rank linear STBC can be full, i.e., $n_t$ for $n_t$ transmit antennas. It has been recently shown that its rate is upper bounded by 1 if a code achieves full diversity with a linear receiver. The intermediate criterion proposed in this paper provides the possibility for codes of rates between $n_t$ and 1 that achieve full diversity with a PIC group decoding. This therefore provides a complexity-performance-rate tradeoff. Some design examples are given.

The authors are with the Department of Electrical and Computer Engineering, University of Delaware, Newark, DE 19716, USA (e-mail: {guo,xxia}@ee.udel.edu). This work was supported in part by the Air Force Office of Scientific Research (AFOSR) under Grant No. FA9550-08-1-0219 and the National Science Foundation under Grant CCR-0325180.

**Index Terms**

## I. Introduction

MIMO technology is an important advancement in wireless communications since it offers significant increase in channel capacity and communication reliability without requiring additional bandwidth or transmission power. Space-time coding is an effective way to explore the promising potential of an MIMO system. In the coherent scenario, where the channel state information (CSI) is available at the receiver, the *full rank* design criterion is derived in [13], [37] to achieve the maximum diversity order in a quasi-static Rayleigh fading channel. However, the derivation of the full rank criterion is based on the assumption of the optimal decoding at the receiver. In order to achieve the maximum diversity order, received signals must be decoded using the maximum likelihood (ML) decoding. Unfortunately, the computational complexity of the ML decoding grows exponentially with the number of the embedded information symbols in the codeword. This often makes the ML decoding infeasible for codes with many information symbols embedded in. Although near-optimal decoding algorithms, such as sphere decoding or lattice-reduction-aided sphere decoding, exist in the literature, [4], [5], [26], [27], [42], their complexities may depend on a channel condition.

In order to significantly reduce the decoding complexity, one may decode one symbol at a time and make the decoding complexity grow linearly with the number of the embedded information symbols. This can be achieved by passing the received signals through a linear filter, which strengthens a main symbol and suppresses all the other interference symbols and then one decodes the main symbol from the output of the filter. By passing the received signal through a filter bank, one can decode each symbol separately. There are different criteria to strengthen the main symbol and suppress the interference symbols. If the filter is designed to completely eliminate the interferences from the other symbols, we call such decoding method *zero-forcing* (ZF) or *interference nulling* decoding. If the filter is designed according to the minimum mean square error (MMSE) criterion, we call the decoding method *MMSE* decoding. The well known algorithms with the above idea are BLAST-SIC algorithms [45]. Since these symbol-by-symbol decoding methods may not be ML but only suboptimal, the *full rank* criterion can not guarantee the codes to achieve the maximum diversity order. In some special cases, the symbol-by-symbol decoding is equivalent to the ML decoding and thus the full rank property ensures the codes achieve full diversity in these cases. The first such a code is the Alamouti code for two transmit antennas [1]. The orthogonal structure of the Alamouti code ensures that symbol-by-symbol decoding is equivalent to the ML decoding. The Alamouti code has inspired many studies on orthogonal STBC (OSTBC) [22], [23], [25], [36], [38], [44]. However, OSTBC suffers from a low symbol rate. In [44], it has been proved that the symbol rate of an OSTBC is upper-bounded by $3/4$ with or without linear processing among the embedded information symbols or their complex conjugates for more than $2$ transmit antennas and conjectured that it is upper bounded by $\frac{k+1}{2k}$ for $2k-1$ and $2k$ transmit antennas, where $k$ is a positive integer (this upper bound was shown in [22] when no linear processing is used among information symbols). Explicit designs of OSTBCs with rates achieving the conjectured upper bound have been given in [22], [25], [36]. Note that the rate only approaches to $1/2$ when the number of transmit antennas goes large. For a general linear dispersion STBC [14], [15] that has no orthogonal structure, the full diversity criterion for STBC decoded with a symbol-by-symbol decoding method has not been discovered until recently. In [48], Zhang-Liu-Wong proposed a family of STBC called *Toeplitz codes* and proved that a Toeplitz code achieves full diversity with the ZF receiver. The symbol rate of a Toeplitz code approaches $1$ as the block length goes to infinity. Later in [32], Shang-Xia extended the result in [48] and proposed a design criterion for the codes achieving full diversity with ZF and MMSE receivers. They also proposed a new family of STBC called *overlapped Alamouti codes (OAC)*, which has better performance than Toeplitz codes for any number of transmit antennas. The symbol rate of an OAC also approaches to $1$ as the block length goes to infinity. It has been proved in [32] that the symbol rate of an STBC achieving full diversity with a linear receiver is upper-bounded by $1$. Simulation results in [32] show that OAC outperform OSTBC for over $4$ transmit antennas. Note that it is shown in [32] that for any OSTBC, its MMSE/ZF receiver is the same as the ML receiver.

Although OSTBCs can be optimally decoded in a symbol-by-symbol way, the orthogonality condition is too restrictive as we mentioned above. From an information theoretical point of view, this can cause a significant loss of channel capacity [30]. By relaxing the orthogonality condition on the code matrix, quasi-orthogonal STBC (QOSTBC) was introduced by Jafarkhani in [16], Tirkkonen-Boariu-Hottinen in [40] and Papadias-Foschini in [30] to improve the symbol rate at the tradeoff of a higher decoding complexity. The basic idea of QOSTBC is to group the column vectors in the code matrix into pairs and keep the orthogonality among the groups of the column vectors while relax the orthogonality requirement within each group. Because of this partial orthogonality structure, QOSTBC can be (ML) decoded pair-by-pair complex symbols, which has a higher decoding complexity compared to the OSTBC. The original QOSTBCs do not possess the full diversity property. The idea of rotating information symbols in a QOSTBC to achieve full diversity and maintain the complex symbol pair-wise ML decoding has appeared independently in [34], [35], [39], and the optimal rotation angles $\pi/4$ and $\pi/6$ of the above mentioned information symbols for any signal constellations on square lattices and equal-literal triangular lattices, respectively, have been obtained in Su-Xia [35] in the sense that the diversity products (coding gains) are maximized. In [19], [43], [46], the authors further studied QOSTBC with minimum decoding complexity. The underlying constellation is assumed to be rectangular QAM, which can be viewed as two PAM constellations. The minimum decoding complexity means the code can be optimally decoded in a real-pair-wise way. Compared to the complex-pair-wise decodable QOSTBC, the decoding complexity of real-pair-wise decodable QOSTBC is lower. In [7], [17], [18], [21], [43], [47], the pair-by-pair decoding was generalized to a general group-by-group decoding. The symbols in a code matrix are separated into several groups and each group is decoded separately. With the help of graph theory, a rate $\frac{5}{4}$ code was obtained in [47] that can be decoded in two groups, each group contains $5$ real symbols. In [17], [18], a Clifford algebra approach is applied for multi-group decodable STBCs.

In this paper, we propose a general decoding scheme called partial interference cancellation (PIC) group decoding algorithm for linear dispersion (complex conjugated symbols may be embedded) space-time block codes (STBC) [14], [15]. A PIC group decoding decodes the symbols embedded in an STBC by dividing them into several groups and decoding each group separately after a linear PIC operation is implemented. It can be viewed as an intermediate decoding between the ML receiver that decodes all the embedded symbols together, i.e., all embedded symbols are in a single group, and the ZF receiver that decodes all the embedded symbols separately and independently, i.e., each group has and only has one embedded symbol, after the ZF operation is implemented. The PIC group decoding provides a framework to adjust the complexity-performance tradeoff by choosing the sizes of the information symbol groups. It contains the previously studied decoding algorithms for codes such as OSTBC [1], [38],

QOSTBC [16], [19], [34], [35], [40], [43], [46], and STBC achieving full diversity with linear receivers [32], [48] as special cases. We propose a design criterion for STBC achieving full diversity with the PIC decoding algorithm. Our proposed design criterion is an intermediate criterion between the loosest ML full rank criterion [13], [37] of codewords and the strongest ZF linear independence criterion of the column vectors in the equivalent channel matrix [32]. We then propose asymptotic optimal (AO) group decoding algorithm which is an intermediate decoding between the MMSE decoding algorithm and the ML decoding algorithm. The design criterion for the PIC group decoding can be applied to the AO group decoding algorithm because of its asymptotic optimality. It is well-known that the symbol rate for a full rank linear STBC can be full, i.e., $n_t$ for $n_t$ transmit antennas. It has been recently shown in [32] that its rate is upper bounded by $1$ if a code achieves full diversity with a linear receiver. The intermediate criterion proposed in this paper provides the possibility for codes of rates between $n_t$ and $1$ that achieve full diversity with the PIC group decoding. This therefore provides a complexity-performance-rate tradeoff. Design examples of STBC achieving full diversity with the PIC group decoding are finally presented. Our simulations show that these codes can perform better than the Alamouti code for 2 transmit antennas and the QOSTBC with the optimal rotation for 4 transmit antennas. Note that a similar algorithm and an STBC design have been proposed lately in [28] but they do not achieve full diversity.

This paper is organized as follows. In Section II, we describe the system model; in Section III, we propose the PIC group decoding algorithm, its connection with ZF decoding algorithm and the corresponding successive interference cancellation (SIC) aided decoding algorithm or PIC-SIC; In Section IV, we systematically study the diversity property of the codes decoded with the PIC group decoding and the PIC-SIC group decoding, and derive the design criterion. In Section V, we propose AO group decoding. In Section VI, we present two design examples. In Section VII, we present some simulation results.

Some notations in this paper are defined as follows.

- $\mathbb{C}$: complex number field;
- $\mathbb{R}$: real number field;
- $\mathcal{A}$: a signal constellation;
- tr: trace of a matrix;
- Bold faced upper-case letters such as $\boldsymbol{A}$ represent matrices;

- Bold faced lower-case letters such as $\boldsymbol{x}$ represent column vectors;
- Superscripts $^\mathsf{T}$, $^\mathsf{H}$, $^*$: transpose, complex conjugate transpose, complex conjugate, respectively;
- $\|\cdot\|$: $l_2$-norm for a vector;
- $\|\cdot\|_F$: Frobenius norm for a matrix;
- $\boldsymbol{i}$: $\sqrt{-1}$.

## II. System Model

We consider a quasi-static Rayleigh block-fading channel with coherence time $t$. Assume there are $n_t$ transmit and $n_r$ receive antennas. The channel model is written as follows,

$$\boldsymbol{Y} = \sqrt{\frac{\mathsf{SNR}}{n_t}}\boldsymbol{H}\boldsymbol{X} + \boldsymbol{W}, \tag{1}$$

where $\boldsymbol{Y} = (y_{i,j}) \in \mathbb{C}^{n_r \times t}$ is the received signal matrix that is received in $t$ time slots, $\boldsymbol{H} = (h_{i,j}) \in \mathbb{C}^{n_r \times n_t}$ is the channel matrix, the entries of $\boldsymbol{H}$ are assumed i.i.d. with distribution $\mathcal{CN}(0,1)$, $\boldsymbol{X} \in \mathbb{C}^{n_t \times t}$ is the codeword matrix that is normalized so that its average energy is $tn_t$, i.e.,

$$\mathrm{tr}\left(\mathcal{E}\left\{\boldsymbol{X}^\mathsf{H}\boldsymbol{X}\right\}\right) = tn_t,$$

$\boldsymbol{W} \in \mathbb{C}^{n_r \times t}$ is the additive white Gaussian noise matrix with i.i.d. entries $w_{i,j} \sim \mathcal{CN}(0,1)$, $\mathsf{SNR}$ is the average signal-to-noise-ratio (SNR) at the receiver.

In this paper, we only consider linear dispersion STBC, which covers most existing STBCs, [14]:

$$\boldsymbol{X} = \sum_{i=0}^{n-1} x_i \boldsymbol{A}_i + x_i^* \boldsymbol{B}_i, \tag{2}$$

where $x_i \in \mathcal{A}, i = 0, 1, \ldots, n-1$, are the embedded information symbols, $\mathcal{A}$ is a signal constellation, $\boldsymbol{A}_i, \boldsymbol{B}_i \in \mathbb{C}^{n_t \times t}, i = 0, 1, \ldots, n-1$, are constant matrices called dispersion matrices. We use $\mathcal{X}$ to denote the codebook, i.e.,

$$\mathcal{X} = \left\{\boldsymbol{X} = \sum_{i=0}^{n-1} x_i \boldsymbol{A}_i + x_i^* \boldsymbol{B}_i, x_i \in \mathcal{A}, i = 0, 1, \ldots, n-1\right\}. \tag{3}$$

For convenience, we also use $\mathcal{X}$ to denote the coding scheme that is associated with the codebook.

In order to apply a linear operation, the system model in (1) needs to be rewritten as

$$\boldsymbol{y} = \sqrt{\mathsf{SNR}}\boldsymbol{G}\boldsymbol{x} + \boldsymbol{w}, \tag{4}$$

where $\boldsymbol{y} \in \mathbb{C}^{tn_r}$ is the received signal vector, $\boldsymbol{G} \in \mathbb{C}^{tn_r \times n}$ is an *equivalent channel matrix* [14], [15], [32]; $\boldsymbol{x} = [x_0, x_1, \ldots, x_{n-1}]^\mathsf{T} \in \mathcal{A}^n$ is the information symbol vector; $\boldsymbol{w} = [w_0, w_1, \ldots, w_{tn_r}]^\mathsf{T} \in \mathbb{C}^{tn_r}$ is the additive white Gaussian noise, $w_i \sim \mathcal{CN}(0,1)$. For many (if not all) existing linear dispersion (or linear lattice) STBCs, such as those in [1], [2], [9]–[11], [15], [16], [20], [29], [31], [32], [35], the channel model can be rewritten in the form of (4). One simple observation is that for a linear dispersion STBC that is defined as

$$\boldsymbol{X} = \sum_{i=0}^{n-1} x_i \boldsymbol{A}_i, \tag{5}$$

which is a special case of the linear dispersion STBC in (2), the channel model can always be written in the form of (4). All the codes in [2], [9]–[11], [15], [20], [29], [31] fall into this category. Another case in which the channel model can be rewritten in the form of (4) is that each column of $\boldsymbol{X}$ contains linear combinations of either only $x_i, i = 0, 1, \ldots, n-1$ or only $x_i^*, i = 0, 1, \ldots, n-1$. Examples of such codes include the Alamouti code [1] and QOSTBCs [16], [35] and OAC [32]. For instance, the channel model of the Alamouti code with one receive antenna is

$$\begin{bmatrix} y_{0,0} & y_{0,1} \end{bmatrix} = \sqrt{\frac{\mathsf{SNR}}{2}} \begin{bmatrix} h_{0,0} & h_{0,1} \end{bmatrix} \begin{bmatrix} x_0 & -x_1^* \\ x_1 & x_0^* \end{bmatrix} + \begin{bmatrix} w_{0,0} & w_{0,1} \end{bmatrix}.$$

By taking unitary linear operations and conjugations, which do not change the probabilistic property of the white Gaussian noise, we can extract the embedded information symbol vector and rewrite the above channel model as follows,

$$\begin{bmatrix} y_{0,0} \\ y_{0,1}^* \end{bmatrix} = \sqrt{\mathsf{SNR}} \left( \frac{1}{\sqrt{2}} \begin{bmatrix} h_{0,0} & h_{0,1} \\ h_{0,1}^* & -h_{0,0}^* \end{bmatrix} \right) \begin{bmatrix} x_0 \\ x_1 \end{bmatrix} + \begin{bmatrix} w_{0,0} \\ w_{0,1}^* \end{bmatrix}. \tag{6}$$

It is shown in [32] that for any OSTBC (a column may include both $x_i$ and $x_j^*$ simultaneously), its equivalent channel (4) exists. In the case when there are multiple receive antennas, an equivalent channel matrix can be derived by noting that at each receiver antenna, the received signal model is of the same form as in (6). For example, if there are two receive antennas for the Alamouti code, then an equivalent channel model is

$$\begin{bmatrix} y_{0,0} \\ y_{0,1}^* \\ y_{1,0} \\ y_{1,1}^* \end{bmatrix} = \sqrt{\mathsf{SNR}} \left( \frac{1}{\sqrt{2}} \begin{bmatrix} h_{0,0} & h_{0,1} \\ h_{0,1}^* & -h_{0,0}^* \\ h_{1,0} & h_{1,1} \\ h_{1,1}^* & -h_{1,0}^* \end{bmatrix} \right) \begin{bmatrix} x_0 \\ x_1 \end{bmatrix} + \begin{bmatrix} w_{0,0} \\ w_{0,1}^* \\ w_{1,0} \\ w_{1,1}^* \end{bmatrix}.$$

It is not hard to see that the original channel $\boldsymbol{H}$ and an equivalent channel $\boldsymbol{G}$ satisfy the following property,

$$\left\|\boldsymbol{H}\left(\boldsymbol{X}_1 - \boldsymbol{X}_2\right)\right\|_F = \left\|\boldsymbol{G}\left(\boldsymbol{x}_1 - \boldsymbol{x}_2\right)\right\|, \tag{7}$$

where $\boldsymbol{X}_1, \boldsymbol{X}_2 \in \mathcal{X}$, $\boldsymbol{x}_1$ and $\boldsymbol{x}_2$ are vectors of information symbols embedded in $\boldsymbol{X}_1$ and $\boldsymbol{X}_2$, respectively.

For a linear dispersion code with a rectangular signal constellation $\mathcal{A}$, which can be viewed as two PAM constellations, if it does not have its equivalent channel model in (4), the channel model can always be written in the following form [6], [14],

$$\boldsymbol{y} = \sqrt{\mathsf{SNR}}\,\boldsymbol{G}\begin{bmatrix}\mathfrak{Re}(\boldsymbol{x})\\ \mathfrak{Im}(\boldsymbol{x})\end{bmatrix} + \boldsymbol{w}, \tag{8}$$

where $\boldsymbol{y} \in \mathbb{R}^{2tn_r}$ is the received signal vector, $\boldsymbol{G} \in \mathbb{R}^{2tn_r \times 2n}$ is the *equivalent channel matrix*, $\boldsymbol{w} = [w_0, w_1, \ldots, w_{2tn_r}]^{\mathsf{T}} \in \mathbb{R}^{2tn_r}$ is the real white Gaussian noise vector, $w_i \sim \mathcal{N}(0, \frac{1}{2})$. The entries of

$$\begin{bmatrix}\mathfrak{Re}(\boldsymbol{x})\\ \mathfrak{Im}(\boldsymbol{x})\end{bmatrix}$$

can be viewed as drawn from a PAM constellation. Hence there is no essential difference between the models in (4) and (8) except that the noise in (8) is real.

Note that for both channel models in (4) and (8), the entries of the equivalent channel matrix $\boldsymbol{G}$ are linear combinations of $h_{i,j}$ and $h^*_{i,j}, 0 \le i \le n_r - 1, 0 \le j \le n_t - 1$. If we use the notation $\boldsymbol{h} = [h_0, h_1, \ldots, h_{l-1}] \triangleq \mathrm{vec}(\boldsymbol{H})$, then both (4) and (8) are special cases of the following model,

$$\boldsymbol{y} = \sqrt{\mathsf{SNR}}\,\boldsymbol{G}(\boldsymbol{h})\boldsymbol{x} + \boldsymbol{w}, \tag{9}$$

where $\boldsymbol{G}(\boldsymbol{h}) \in \mathbb{C}^{m \times n}$ is an equivalent channel matrix, which is a function of $\boldsymbol{h} = [h_0, h_1, \ldots, h_{l-1}]$, $h_i \sim \mathcal{CN}(0, 1)$, $\boldsymbol{x} = [x_0, x_1, \ldots, x_{n-1}] \in \mathcal{A}^n$ is the information symbol vector, $\boldsymbol{w} = [w_0, \ldots, w_{m-1}]$ is the additive white Gaussian noise vector. For convenience, we always assume that noise $\boldsymbol{w}$ is complex Gaussian, while for real Gaussian $\boldsymbol{w}$, the derivation is exactly the same. From the following discussions, we shall see later that not only the channel model in (9) contains the equivalent channel model derived from transforming the original channel model of linear dispersion STBC in (1), but also it is a resulted form after each PIC operation.

## III. PIC Group Decoding Algorithm

In this section, we present a PIC group decoding algorithm that is, as we mentioned before, an intermediate decoding algorithm between the ML decoding algorithm and the ZF decoding algorithm, and has the ML decoding and the ZF decoding as two special cases. In the first subsection, we describe the PIC group decoding algorithm; in the second subsection, we discuss its connection with the ZF decoding algorithm; in the third subsection, we discuss the successive interference cancellation aided PIC group decoding algorithm (PIC-SIC); some examples are given in the last part of this section to illustrate the PIC group decoding algorithm.

### A. *Partial Interference Cancellation Group Decoding Algorithm*

We now present a detailed description of the PIC group decoding algorithm. All the following discussions are based on the equivalent channel model in (9). First let us introduce some notations. Define index set $\mathcal{I}$ as

$$\mathcal{I} = \{0, 1, 2, \ldots, n-1\},$$

where $n$ is the number of information symbols in $\boldsymbol{x}$. First we partition $\mathcal{I}$ into $N$ groups: $\mathcal{I}_0, \mathcal{I}_1, \ldots, \mathcal{I}_{N-1}$. Each index subset $\mathcal{I}_k$ can be written as follows,

$$\mathcal{I}_k = \{i_{k,0}, i_{k,1}, \ldots, i_{k,n_k-1}\}, \quad k = 0, 1, \ldots, N-1,$$

where $n_k \triangleq |\mathcal{I}_k|$ is the cardinality of the subset $\mathcal{I}_k$. We call $\mathcal{I} = \{\mathcal{I}_0, \mathcal{I}_1, \ldots, \mathcal{I}_{N-1}\}$ a grouping scheme, where, for simplicity, we still use $\mathcal{I}$ to denote a grouping scheme. For such a grouping scheme, the following two equations must hold,

$$\mathcal{I} = \bigcup_{i=0}^{N-1} \mathcal{I}_i \text{ and } \sum_{i=0}^{N-1} n_i = n.$$

Define $\boldsymbol{x}_{\mathcal{I}_k}$ as the information symbol vector that contains the symbols with indices in $\mathcal{I}_k$, i.e.,

$$\boldsymbol{x}_{\mathcal{I}_k} = \left[x_{i_{k,0}}, x_{i_{k,1}}, \ldots, x_{i_{k,n_k-1}}\right]^{\mathsf{T}}.$$

Let the column vectors of an equivalent channel matrix $\boldsymbol{G}(\boldsymbol{h})$ be $\boldsymbol{g}_0, \boldsymbol{g}_1, \ldots, \boldsymbol{g}_{n-1}$ that have size $m \times 1$. Then, we can similarly define $\boldsymbol{G}_{\mathcal{I}_k}$ as

$$\boldsymbol{G}_{\mathcal{I}_k} = \left[\boldsymbol{g}_{i_{k,0}}, \boldsymbol{g}_{i_{k,1}}, \ldots, \boldsymbol{g}_{i_{k,n_k-1}}\right]. \tag{10}$$

With these notations, equation (9) can be written as

$$\boldsymbol{y} = \sqrt{\mathsf{SNR}} \sum_{i=0}^{N-1} \boldsymbol{G}_{\mathcal{I}_i} \boldsymbol{x}_{\mathcal{I}_i} + \boldsymbol{w}. \tag{11}$$

Suppose we want to decode the $k$-th symbol group $\boldsymbol{x}_{\mathcal{I}_k}$. Note that in the ZF decoding algorithm, to decode the $k$-th symbol, the interferences from the other symbols are completely eliminated by a linear filter (the $k$-th row of the pseudo-inverse matrix of the equivalent channel). The same idea can be applied here. We want to find a matrix (linear filter) $\boldsymbol{P}_{\mathcal{I}_k}$ such that by multiplying $\boldsymbol{y}$ by $\boldsymbol{P}_{\mathcal{I}_k}$ to the left (linear filtering), all the interferences from the other symbol groups can be eliminated. Such a matrix $\boldsymbol{P}_{\mathcal{I}_k}$ can be found as follows. Define $\boldsymbol{Q}_{\mathcal{I}_k} \in \mathbb{C}^{m \times m}$ as the projection matrix that projects a vector in $\mathbb{C}^m$ to the subspace $\mathbb{V}_{\mathcal{I}_k}$ that is defined as

$$\mathbb{V}_{\mathcal{I}_k} = \operatorname{span}\left\{\boldsymbol{g}_i, 0 \le i < n, i \notin \mathcal{I}_k\right\}. \tag{12}$$

Let $\boldsymbol{G}^c_{\mathcal{I}_k} \in \mathbb{C}^{m \times (n-n_k)}$ denote the matrix that is obtained by removing the column vectors in $\boldsymbol{G}$ with indices in $\mathcal{I}_k$, i.e.,

$$\boldsymbol{G}^c_{\mathcal{I}_k} = \left[\boldsymbol{G}_{\mathcal{I}_0}, \boldsymbol{G}_{\mathcal{I}_1}, \ldots, \boldsymbol{G}_{\mathcal{I}_{k-1}}, \boldsymbol{G}_{\mathcal{I}_{k+1}}, \ldots, \boldsymbol{G}_{\mathcal{I}_{N-1}}\right]. \tag{13}$$

Then, the projection matrix $\boldsymbol{Q}_{\mathcal{I}_k}$ can be expressed in terms of $\boldsymbol{G}^c_{\mathcal{I}_k}$ as follows,

$$\boldsymbol{Q}_{\mathcal{I}_k} = \boldsymbol{G}^c_{\mathcal{I}_k} \left(\left(\boldsymbol{G}^c_{\mathcal{I}_k}\right)^{\mathsf{H}} \boldsymbol{G}^c_{\mathcal{I}_k}\right)^{-1} \left(\boldsymbol{G}^c_{\mathcal{I}_k}\right)^{\mathsf{H}}, \tag{14}$$

where we assume $\boldsymbol{G}^c_{\mathcal{I}_k}$ is full column rank. If $\boldsymbol{G}^c_{\mathcal{I}_k}$ is not full column rank, then we need to pick a maximal linear independent vector group from $\boldsymbol{G}^c_{\mathcal{I}_k}$ and furthermore, in this case the symbols in group $\mathcal{I}_k$ can not be solved, which is out off the scope of this paper. Thus, we always assume that $\boldsymbol{G}^c_{\mathcal{I}_k}$ has full column rank. Define $\boldsymbol{P}_{\mathcal{I}_k}$ as

$$\boldsymbol{P}_{\mathcal{I}_k} = \boldsymbol{I}_m - \boldsymbol{Q}_{\mathcal{I}_k}, \tag{15}$$

then $\boldsymbol{P}_{\mathcal{I}_k}$ is the projection matrix that projects a vector in $\mathbb{C}^m$ onto the orthogonal complementary subspace $\mathbb{V}^{\perp}_{\mathcal{I}_k}$. Since the projection of any vector in $\mathbb{V}_{\mathcal{I}_k}$ onto $\mathbb{V}^{\perp}_{\mathcal{I}_k}$ is a zero vector, we have

$$\boldsymbol{P}_{\mathcal{I}_k} \boldsymbol{G}_{\mathcal{I}_i} = \boldsymbol{0},\ i = 0, 1, \ldots, k-1, k+1, \ldots, N-1, \tag{16}$$

which is due to $\boldsymbol{P}_{\mathcal{I}_k} \boldsymbol{G}^c_{\mathcal{I}_k} = \boldsymbol{0}$. Define $\boldsymbol{z}_{\mathcal{I}_k} \triangleq \boldsymbol{P}_{\mathcal{I}_k} \boldsymbol{y}$. By applying (16), we get

$$\begin{aligned} \boldsymbol{z}_{\mathcal{I}_k} &= \sqrt{\mathsf{SNR}} \boldsymbol{P}_{\mathcal{I}_k} \sum_{i=0}^{N-1} \boldsymbol{G}_{\mathcal{I}_i} \boldsymbol{x}_{\mathcal{I}_i} + \boldsymbol{P}_{\mathcal{I}_k} \boldsymbol{w} \\ &= \sqrt{\mathsf{SNR}} \boldsymbol{P}_{\mathcal{I}_k} \boldsymbol{G}_{\mathcal{I}_k} \boldsymbol{x}_{\mathcal{I}_k} + \boldsymbol{P}_{\mathcal{I}_k} \boldsymbol{w}. \end{aligned} \tag{17}$$

From (17), we can see that by passing the received signal vector $\boldsymbol{y}$ through the linear filter $\boldsymbol{P}_{\mathcal{I}_k}$, the interferences from the other symbol groups are completely canceled and the output $\boldsymbol{z}_{\mathcal{I}_k}$ only contains the components of the symbol group $\boldsymbol{x}_{\mathcal{I}_k}$. There may exist other matrices that can remove the components of the interference symbol groups in $\boldsymbol{y}$. The following lemma shows that the linear filter matrix $\boldsymbol{P}_{\mathcal{I}_k}$ defined above is the best choice.

**Lemma 1.** *Consider the channel model in* (11) *and let* $\mathsf{SNR}$ *be the SNR of the system. Suppose we want to detect the symbol group* $\boldsymbol{x}_{\mathcal{I}_k}$*. Let* $\mathcal{P}_{\mathcal{I}_k}$ *be the matrix set that contains all the matrices that can cancel the interferences from* $\boldsymbol{x}_{\mathcal{I}_i}, 0 \le i < N, i \ne k$*, i.e.,*

$$\mathcal{P}_{\mathcal{I}_k} = \left\{ \tilde{\boldsymbol{P}}_{\mathcal{I}_k} \,\middle|\, \tilde{\boldsymbol{P}}_{\mathcal{I}_k} \boldsymbol{G}_{\mathcal{I}_i} = \boldsymbol{0},\ 0 \le i < N, i \ne k \right\}. \tag{18}$$

*The block error probability of the system*

$$\tilde{\boldsymbol{z}}_{\mathcal{I}_k} \triangleq \tilde{\boldsymbol{P}}_{\mathcal{I}_k} \boldsymbol{y} = \sqrt{\mathsf{SNR}} \tilde{\boldsymbol{P}}_{\mathcal{I}_k} \boldsymbol{G}_{\mathcal{I}_k} \boldsymbol{x}_{\mathcal{I}_k} + \tilde{\boldsymbol{P}}_{\mathcal{I}_k} \boldsymbol{w} \tag{19}$$

*from ML decoding is denoted as* $P_{\mathsf{err}}(\tilde{\boldsymbol{P}}_{\mathcal{I}_k}, \mathsf{SNR})$*. Then for any given* $\mathsf{SNR}$*, we always have*

$$\boldsymbol{P}_{\mathcal{I}_k} = \arg \min_{\tilde{\boldsymbol{P}}_{\mathcal{I}_k} \in \mathcal{P}_{\mathcal{I}_k}} P_{\mathsf{err}}(\tilde{\boldsymbol{P}}_{\mathcal{I}_k}, \mathsf{SNR}),$$

*where* $\boldsymbol{P}_{\mathcal{I}_k}$ *is defined as in* (15).

A proof of this lemma is given in Appendix A.

Notice that since in our PIC group decoding, all the symbols in a group are decoded together, using highest SNR as the optimality may not be proper. This is the reason why in the above lemma, block error probability is used as the criterion for the optimality of a filter. Equation (17) can be viewed as a channel model in which $\boldsymbol{x}_{\mathcal{I}_k}$ is the transmitted signal vector and $\boldsymbol{z}_{\mathcal{I}_k}$ is the received signal vector. As we mentioned before in Section II, this channel model is derived from the interference cancellation procedure, and fits into the general channel model in (9). Note that in (17), the noise term $\boldsymbol{P}_{\mathcal{I}_k} \boldsymbol{w}$ is no longer a white Gaussian noise. Despite of the presence of this non-white Gaussian noise term, the following lemma shows that the minimum distance decision is still the ML decision in this case.

**Lemma 2.** *Consider the channel model*

$$\boldsymbol{y} = \sqrt{\mathsf{SNR}} \boldsymbol{G} \boldsymbol{x} + \tilde{\boldsymbol{w}}, \tag{20}$$

*where $\boldsymbol{G} \in \mathbb{C}^{m\times n}$ is the channel matrix that is known at the receiver, $\boldsymbol{x} \in \mathcal{A}^n$ is the information symbol vector, $\tilde{\boldsymbol{w}} = \boldsymbol{P}\boldsymbol{w} \in \mathbb{C}^m$, $\boldsymbol{w}$ is the white Gaussian noise vector and $\boldsymbol{P} \in \mathbb{C}^{m\times m}$ is a projection matrix that projects a vector in $\mathbb{C}^m$ to a subspace $\mathbb{V} \subset \mathbb{C}^m$. Assume the column vectors of $\boldsymbol{G}$ belong to $\mathbb{V}$. Then, the decision made by*

$$\hat{\boldsymbol{x}} = \arg\min_{\bar{\boldsymbol{x}}\in\mathcal{A}^n} \left\|\boldsymbol{y} - \sqrt{\mathsf{SNR}}\boldsymbol{G}\bar{\boldsymbol{x}}\right\|$$

*is the ML decision.*

An intuitive explanation for the above lemma is that $\tilde{\boldsymbol{w}}$ is a degenerated white Gaussian noise, which can be a white Gaussian noise by removing some extra dimensions. Its detailed proof is given in Appendix B. According to Lemma 2, the optimal detection of $\boldsymbol{x}_{\mathcal{I}_k}$ from $\boldsymbol{z}_{\mathcal{I}_k}$ is made by

$$\hat{\boldsymbol{x}}_{\mathcal{I}_k} = \arg\min_{\bar{\boldsymbol{x}}\in\mathcal{A}^{n_k}} \left\|\boldsymbol{z}_{\mathcal{I}_k} - \sqrt{\mathsf{SNR}}\boldsymbol{P}_{\mathcal{I}_k}\boldsymbol{G}_{\mathcal{I}_k}\bar{\boldsymbol{x}}\right\|, \tag{21}$$

which is the *PIC group decoding algorithm* we propose in this paper. The complexity of the ML decoding of the dimension-reduced system in (17) is obviously lower than that of the original system in (11). The PIC group decoding algorithm (21) can be viewed as a decomposition of the original high-dimensional decoding problem with high complexity into low-dimensional decoding problem with relatively low decoding complexity. In the extreme case when all the symbols are grouped together, i.e., the problem is not decomposed at all, the PIC group decoding is the same as the ML decoding. In another extreme case when each symbol forms a group, i.e., the problem is completely decomposed, the PIC group decoding is equivalent to the ZF decoding. The detailed description of the connection between these two is given in the following subsection.

### *B. Connection Between PIC Group Decoding and ZF Decoding*

In this subsection we discuss the connection between the PIC group decoding algorithm and the ZF decoding algorithm. In the case when the decoding problem is completely decomposed, i.e., each symbol group contains only one symbol, the PIC group decoding algorithm becomes a symbol-by-symbol decoding algorithm. The grouping scheme in this case is

$$\mathcal{I} = \{\mathcal{I}_0, \mathcal{I}_1, \ldots, \mathcal{I}_{n-1}\} = \{\{0\}, \{1\}, \ldots, \{n-1\}\},$$

where $\mathcal{I}_i = \{i\}$, $0 \le i \le n-1$. To simplify the notations, we use $i$ to denote $\mathcal{I}_i$ in this special case, hence $\boldsymbol{P}_i = \boldsymbol{P}_{\mathcal{I}_i}$. All the other notations with $\mathcal{I}_i$ as the subscript can be similarly defined.

Furthermore, we have $x_i = \boldsymbol{x}_i = \boldsymbol{x}_{\mathcal{I}_i}$, $\boldsymbol{g}_i = \boldsymbol{G}_i = \boldsymbol{G}_{\mathcal{I}_i}$. From the PIC group decoding algorithm (21), the ML decision of $x_k$ from $\boldsymbol{z}_k$ is made by

$$\hat{x}_k = \arg\min_{\bar{x}_k \in \mathcal{A}} \left\| \boldsymbol{z}_k - \sqrt{\mathsf{SNR}} \boldsymbol{P}_k \boldsymbol{g}_k \bar{x}_k \right\|^2 \tag{22}$$

$$= \arg\min_{\bar{x}_k \in \mathcal{A}} \left\| \frac{(\boldsymbol{P}_k \boldsymbol{g}_k)(\boldsymbol{P}_k \boldsymbol{g}_k)^{\mathsf{H}}}{\|\boldsymbol{P}_k \boldsymbol{g}_k\|^2} \boldsymbol{z}_k - \sqrt{\mathsf{SNR}} \boldsymbol{P}_k \boldsymbol{g}_k \bar{x}_k + \left( \boldsymbol{I} - \frac{(\boldsymbol{P}_k \boldsymbol{g}_k)(\boldsymbol{P}_k \boldsymbol{g}_k)^{\mathsf{H}}}{\|\boldsymbol{P}_k \boldsymbol{g}_k\|^2} \right) \boldsymbol{z}_k \right\|^2 . \tag{23}$$

It is easy to verify that in (23) the last term is orthogonal to the first two terms, thus the above decision-making rule can be rewritten as

$$\begin{aligned}
\hat{x}_k &= \arg\min_{\bar{x}_k \in \mathcal{A}} \left( \left\| \frac{(\boldsymbol{P}_k \boldsymbol{g}_k)(\boldsymbol{P}_k \boldsymbol{g}_k)^{\mathsf{H}}}{\|\boldsymbol{P}_k \boldsymbol{g}_k\|^2} \boldsymbol{z}_k - \sqrt{\mathsf{SNR}} \boldsymbol{P}_k \boldsymbol{g}_k \bar{x}_k \right\|^2 + \left\| \left( \boldsymbol{I} - \frac{(\boldsymbol{P}_k \boldsymbol{g}_k)(\boldsymbol{P}_k \boldsymbol{g}_k)^{\mathsf{H}}}{\|\boldsymbol{P}_k \boldsymbol{g}_k\|^2} \right) \boldsymbol{z}_k \right\|^2 \right) \\
&= \arg\min_{\bar{x}_k \in \mathcal{A}} \left\| \sqrt{\mathsf{SNR}} \boldsymbol{P}_k \boldsymbol{g}_k \left( \frac{(\boldsymbol{P}_k \boldsymbol{g}_k)^{\mathsf{H}}}{\sqrt{\mathsf{SNR}} \|\boldsymbol{P}_k \boldsymbol{g}_k\|^2} \boldsymbol{z}_k - \bar{x}_k \right) \right\|^2 \\
&= \arg\min_{\bar{x}_k \in \mathcal{A}} \left| \frac{(\boldsymbol{P}_k \boldsymbol{g}_k)^{\mathsf{H}}}{\sqrt{\mathsf{SNR}} \|\boldsymbol{P}_k \boldsymbol{g}_k\|^2} \boldsymbol{z}_k - \bar{x}_k \right| \\
&= \arg\min_{\bar{x}_k \in \mathcal{A}} \left| \frac{(\boldsymbol{P}_k \boldsymbol{g}_k)^{\mathsf{H}} \boldsymbol{P}_k}{\sqrt{\mathsf{SNR}} \|\boldsymbol{P}_k \boldsymbol{g}_k\|^2} \boldsymbol{y} - \bar{x}_k \right| \\
&= \arg\min_{\bar{x}_k \in \mathcal{A}} \left| \frac{(\boldsymbol{P}_k \boldsymbol{g}_k)^{\mathsf{H}}}{\sqrt{\mathsf{SNR}} \|\boldsymbol{P}_k \boldsymbol{g}_k\|^2} \boldsymbol{y} - \bar{x}_k \right| .
\end{aligned}$$

The last equality is the result of the Hermitian property and the idempotent property of the projection matrix, i.e., $\boldsymbol{P}_k^{\mathsf{H}} = \boldsymbol{P}_k$ and $\boldsymbol{P}_k^2 = \boldsymbol{P}_k$. Note that

$$\left( \frac{(\boldsymbol{P}_k \boldsymbol{g}_k)^{\mathsf{H}}}{\|\boldsymbol{P}_k \boldsymbol{g}_k\|^2} \right) \boldsymbol{g}_k = \frac{\boldsymbol{g}_k^{\mathsf{H}} \boldsymbol{P}_k \boldsymbol{g}_k}{\boldsymbol{g}_k^{\mathsf{H}} \boldsymbol{P}_k \boldsymbol{P}_k^{\mathsf{H}} \boldsymbol{g}_k} = 1,$$

and by applying (16) we have

$$\left( \frac{(\boldsymbol{P}_k \boldsymbol{g}_k)^{\mathsf{H}}}{\|\boldsymbol{P}_k \boldsymbol{g}_k\|^2} \right) \boldsymbol{g}_i = \frac{\boldsymbol{g}_k^{\mathsf{H}} (\boldsymbol{P}_k \boldsymbol{g}_i)}{\boldsymbol{g}_k^{\mathsf{H}} \boldsymbol{P}_k \boldsymbol{P}_k^{\mathsf{H}} \boldsymbol{g}_k} = 0.$$

The above two equations imply that $\frac{(\boldsymbol{P}_k \boldsymbol{g}_k)^{\mathsf{H}}}{\|\boldsymbol{P}_k \boldsymbol{g}_k\|^2}$ is the $k$-th row of $\left(\boldsymbol{G}(\boldsymbol{h})^{\mathsf{H}} \boldsymbol{G}(\boldsymbol{h})\right)^{-1} \boldsymbol{G}(\boldsymbol{h})^{\mathsf{H}}$. Thus, in the complete decomposition case, the PIC group decoding algorithm is equivalent to the ZF decoding algorithm.

One negative effect of the interference cancellation procedure is that it may reduce the power gain of the symbol $x_k$. Before the interference cancellation, the power gain of $x_k$ is $\|\boldsymbol{g}_k\|^2$, while after the interference cancellation, the power gain of $x_k$ becomes $\|\boldsymbol{P}_k\boldsymbol{g}_k\|^2$. Since $\boldsymbol{P}_k$ is a projection matrix, we always have

$$\|\boldsymbol{P}_k\boldsymbol{g}_k\| \leq \|\boldsymbol{g}_k\| \,.$$

The equality holds if and only if $\boldsymbol{g}_k$ is orthogonal to the space spanned by $\boldsymbol{g}_i, i = 0, 1, \ldots, k-1, k+1, \ldots, n-1$. In the case of OSTBC, the columns of the equivalent channel are orthogonal to each other, and therefore, there is no power gain loss during the interference cancellation. Hence the performance of the ZF receiver is the same as the ML receiver for OSTBC. For all non-orthogonal STBC, an interference cancellation algorithm usually causes a power gain loss and therefore performance loss compared to the ML decoding.

### *C. PIC-SIC Group Decoding Algorithm*

Notice that in the ZF decoding algorithm, we may use *successive interference cancellation (SIC)* strategy to aid the decoding process. We call the SIC-aided ZF decoding algorithm *ZF-SIC decoding algorithm* [41], [45]. The basic idea of SIC is simple: remove the already-decoded symbols from the received signals to reduce the interferences. When the SNR is relatively high, the symbol error rate (SER) of the already-decoded symbols is low and there is a considerable SER performance gain by using the SIC strategy. The same strategy can also be used to aid the PIC group decoding process to improve the SER performance. We call the SIC-aided PIC group decoding algorithm *PIC-SIC group decoding algorithm*. In the PIC group decoding algorithm, the decoding order has no effect on the SER performance. For the PIC-SIC group decoding algorithm, different decoding orders will result in different SER performances. We can obtain a better performance by choosing a proper decoding order. One simple way to determine the decoding order is to use maximum SNR as the criterion to arrange the decoding order, similar to BLAST ordered SIC algorithm. Suppose the ordered symbol sets are as follows,

$$\boldsymbol{x}_{\mathcal{I}_{i_0}}, \boldsymbol{x}_{\mathcal{I}_{i_1}}, \ldots, \boldsymbol{x}_{\mathcal{I}_{i_{N-1}}} \,. \tag{24}$$

The ordered PIC-SIC group decoding algorithm is then:

1) Decode the first set of symbols $\boldsymbol{x}_{\mathcal{I}_{i_0}}$ using the PIC group decoding algorithm (21);
2) Let $k = 0$, $\boldsymbol{y}_0 = \boldsymbol{y}$, where $\boldsymbol{y}$ is defined as in (11);

3) Remove the components of the already-detected symbol set $\boldsymbol{x}_{\mathcal{I}_{i_k}}$ from (11),

$$\boldsymbol{y}_{k+1} \triangleq \boldsymbol{y}_k - \sqrt{\mathsf{SNR}}\boldsymbol{G}_{\mathcal{I}_{i_k}}\boldsymbol{x}_{\mathcal{I}_{i_k}} = \sqrt{\mathsf{SNR}}\sum_{j=k+1}^{N-1}\boldsymbol{G}_{\mathcal{I}_{i_j}}\boldsymbol{x}_{\mathcal{I}_{i_j}} + \boldsymbol{w}; \tag{25}$$

4) Decode $\boldsymbol{x}_{\mathcal{I}_{i_{k+1}}}$ in (25) using the PIC group decoding algorithm;

5) If $k < N-1$, then set $k := k+1$, go to Step 3; otherwise stop the algorithm.

**Remark 1.** For the PIC group decoding algorithm, the equivalent channel matrix $\boldsymbol{G}(\boldsymbol{h}) \in \mathbb{C}^{m\times n}$ must satisfy the condition $\mathbb{V}_{\mathcal{I}_k} \subsetneq \mathbb{C}^m$, otherwise $\boldsymbol{z}_{\mathcal{I}_k} = \mathbf{0}$, i.e., there is no information left in $\boldsymbol{z}_{\mathcal{I}_k}$ about $\boldsymbol{x}_{\mathcal{I}_k}$. This requirement is generally weaker than that of the ZF decoding, which requires that $m \geq n$. For example, consider an uncoded MIMO system with $5$ transmit antennas and $4$ receive antennas. In this case, the ZF receiver can not decode the received signals, while the PIC group decoding with the grouping scheme $\mathcal{I} = \{\mathcal{I}_0 = \{0,1,2\}, \mathcal{I}_1 = \{3,4\}\}$ can do the decoding.

**Remark 2.** For the PIC-SIC group decoding algorithm, we require that at each decoding stage, $\mathbb{V}_{\mathcal{I}_k} \subsetneq \mathbb{C}^m$. This requirement is even weaker than that of the PIC group decoding. Since we remove the interferences from the already-decoded symbols, the subspace $\mathbb{V}_{\mathcal{I}_k}$ shrinks each time when we finish decoding one symbol group. Consider the uncoded MIMO system in Remark 1. Let $\mathcal{I} = \{\mathcal{I}_0 = \{0,1,2,3\}, \mathcal{I}_1 = \{4\}\}$ be the grouping scheme. Then it is not possible to decode the second group symbol $x_4$ with the PIC group decoding algorithm, because after we remove the interferences from $x_0, x_1, x_2, x_3$, there is nothing left ($\boldsymbol{z}_{\mathcal{I}_1} = 0$) due to the lack of dimensionality. However, we can decode $x_4$ with the PIC-SIC group decoding.

## D. Examples

Next we give some examples to illustrate the PIC group decoding algorithm.

*1) Example 1:* Consider the Alamouti code with one receive antenna. The equivalent channel matrix can be written as

$$\boldsymbol{G} = [\boldsymbol{g}_0, \boldsymbol{g}_1],$$

where

$$\boldsymbol{g}_0 = \frac{1}{\sqrt{2}}\begin{bmatrix} h_0 \\ h_1^* \end{bmatrix}, \quad \boldsymbol{g}_1 = \frac{1}{\sqrt{2}}\begin{bmatrix} h_1 \\ -h_0^* \end{bmatrix}.$$

The grouping scheme is $\mathcal{I} = \{\mathcal{I}_0, \mathcal{I}_1\} = \{\{0\}, \{1\}\}$. By a direct computation, we get the projection matrix $\boldsymbol{P}_0$ as follows,

$$\boldsymbol{P}_0 = \frac{1}{|h_0|^2 + |h_1|^2} \begin{bmatrix} |h_0|^2 & h_0 h_1 \\ h_0^* h_1^* & |h_1|^2 \end{bmatrix}.$$

Then, the optimal detection of $x_0$ is

$$\begin{aligned} \hat{x}_0 &= \arg\min_{\bar{x}_0 \in \mathcal{A}} \left\| \boldsymbol{P}_0 \boldsymbol{y} - \sqrt{\mathsf{SNR}} \boldsymbol{P}_0 \boldsymbol{g}_0 \bar{x}_0 \right\| \\ &= \arg\min_{\bar{x}_0 \in \mathcal{A}} \left| h_0^* y_0 + h_1 y_1 - \sqrt{\frac{\mathsf{SNR}}{2}} \left( |h_0|^2 + |h_1|^2 \right) \bar{x}_0 \right|, \end{aligned}$$

which is the same as the optimal detection formula derived in [1]. Similarly, the optimal detection for $x_1$ is

$$\hat{x}_1 = \arg\min_{\bar{x}_1 \in \mathcal{A}} \left| h_1^* y_0 - h_0 y_1 - \sqrt{\frac{\mathsf{SNR}}{2}} \left( |h_0|^2 + |h_1|^2 \right) \bar{x}_1 \right|.$$

*2) Example 2:* Consider the quasi-orthogonal STBC proposed in [40]. The code has the following form,

$$\boldsymbol{X} = \begin{bmatrix} x_0 & -x_1^* & x_2 & -x_3^* \\ x_1 & x_0^* & x_3 & x_2^* \\ x_2 & -x_3^* & x_0 & -x_1^* \\ x_3 & x_2^* & x_1 & x_0^* \end{bmatrix}.$$

Suppose we use one receive antenna. The equivalent channel matrix $\boldsymbol{G}(\boldsymbol{h})$ can be written as

$$\boldsymbol{G} = [\boldsymbol{g}_0, \boldsymbol{g}_1, \boldsymbol{g}_2, \boldsymbol{g}_3],$$

where

$$\boldsymbol{g}_0 = \frac{1}{2} \begin{bmatrix} h_0 \\ h_1^* \\ h_2 \\ h_3^* \end{bmatrix}, \ \boldsymbol{g}_1 = \frac{1}{2} \begin{bmatrix} h_1 \\ -h_0^* \\ h_3 \\ -h_2^* \end{bmatrix}, \ \boldsymbol{g}_2 = \frac{1}{2} \begin{bmatrix} h_2 \\ h_3^* \\ h_3 \\ h_1^* \end{bmatrix}, \ \boldsymbol{g}_3 = \frac{1}{2} \begin{bmatrix} h_3 \\ -h_2^* \\ h_1 \\ -h_0^* \end{bmatrix}.$$

Let $\mathcal{I}_0 = \{0, 2\}$ and $\mathcal{I}_1 = \{1, 3\}$. Then, the optimal detection of $\boldsymbol{x}_{\mathcal{I}_0}$ is

$$\hat{\boldsymbol{x}}_{\mathcal{I}_0} = \arg\min_{\bar{\boldsymbol{x}}_{\mathcal{I}_0} \in \mathcal{A}^2} \left\| \boldsymbol{P}_{\mathcal{I}_0} \boldsymbol{y} - \sqrt{\mathsf{SNR}} \boldsymbol{P}_{\mathcal{I}_0} \boldsymbol{G}_{\mathcal{I}_0} \bar{\boldsymbol{x}}_{\mathcal{I}_0} \right\|. \qquad (26)$$

It is easy to verify that

$$\boldsymbol{g}_0 \perp \boldsymbol{g}_1, \ \boldsymbol{g}_0 \perp \boldsymbol{g}_3, \ \boldsymbol{g}_2 \perp \boldsymbol{g}_1, \ \boldsymbol{g}_2 \perp \boldsymbol{g}_3,$$

so $\mathbb{V}_{\mathcal{I}_0} \perp \mathbb{V}_{\mathcal{I}_1}$. This fact implies that $\boldsymbol{P}_{\mathcal{I}_1}\boldsymbol{G}_{\mathcal{I}_0} = 0$. The decoding rule in (26) can be simplified as

$$\begin{aligned}\hat{\boldsymbol{x}}_{\mathcal{I}_0} &= \arg\min_{\bar{\boldsymbol{x}}_{\mathcal{I}_0}\in\mathcal{A}^2}\left(\left\|\boldsymbol{P}_{\mathcal{I}_0}\boldsymbol{y} - \sqrt{\mathsf{SNR}}\boldsymbol{P}_{\mathcal{I}_0}\boldsymbol{G}_{\mathcal{I}_0}\bar{\boldsymbol{x}}_{\mathcal{I}_0}\right\| + \|\boldsymbol{P}_{\mathcal{I}_1}\boldsymbol{y}\|\right)\\ &= \arg\min_{\bar{\boldsymbol{x}}_{\mathcal{I}_0}\in\mathcal{A}^2}\left(\left\|\boldsymbol{P}_{\mathcal{I}_0}\boldsymbol{y} - \sqrt{\mathsf{SNR}}\boldsymbol{P}_{\mathcal{I}_0}\boldsymbol{G}_{\mathcal{I}_0}\bar{\boldsymbol{x}}_{\mathcal{I}_0}\right\| + \left\|\boldsymbol{P}_{\mathcal{I}_1}\boldsymbol{y} - \sqrt{\mathsf{SNR}}\boldsymbol{P}_{\mathcal{I}_1}\boldsymbol{G}_{\mathcal{I}_0}\bar{\boldsymbol{x}}_{\mathcal{I}_0}\right\|\right)\\ &= \arg\min_{\bar{\boldsymbol{x}}_{\mathcal{I}_0}\in\mathcal{A}^2}\left\|\boldsymbol{y} - \sqrt{\mathsf{SNR}}\boldsymbol{G}_{\mathcal{I}_0}\bar{\boldsymbol{x}}_{\mathcal{I}_0}\right\|.\end{aligned}$$

The decoding rule of $\boldsymbol{x}_{\mathcal{I}_1}$ can be similarly derived,

$$\hat{\boldsymbol{x}}_{\mathcal{I}_1} = \arg\min_{\bar{\boldsymbol{x}}_{\mathcal{I}_1}\in\mathcal{A}^2}\left\|\boldsymbol{y} - \sqrt{\mathsf{SNR}}\boldsymbol{G}_{\mathcal{I}_1}\bar{\boldsymbol{x}}_{\mathcal{I}_1}\right\|.$$

From the above equations, we can see that if the groups are orthogonal to each other, then the decomposition of the system is easy: just to pick up the column vectors corresponding to a group in $\boldsymbol{G}(\boldsymbol{h})$ and get a new equivalent channel matrix, then use this new channel matrix and the received signal $\boldsymbol{y}$ to do the ML decoding. In this case, no linear filtering is needed in the PIC group decoding and the ML decoding and the PIC group decoding are the same.

*3) Example 3:* Consider the $3$ by $8$ overlapped Alamouti code in [32],

$$\boldsymbol{X} = \begin{bmatrix} x_0^* & 0 & x_2^* & x_1 & x_4^* & x_3 & 0 & x_5\\ 0 & x_0 & -x_1^* & x_2 & -x_3^* & x_4 & -x_5^* & 0\\ 0 & x_1^* & x_0 & x_3^* & x_2 & x_5^* & x_4 & 0\end{bmatrix}.$$

An equivalent channel matrix can be written as

$$\boldsymbol{G} = \begin{bmatrix}\boldsymbol{g}_0 & \boldsymbol{g}_1 & \boldsymbol{g}_2 & \boldsymbol{g}_3 & \boldsymbol{g}_4 & \boldsymbol{g}_5\end{bmatrix} = \frac{1}{\sqrt{3}}\begin{bmatrix} h_0^* & 0 & 0 & 0 & 0 & 0\\ h_1 & h_2 & 0 & 0 & 0 & 0\\ h_2^* & -h_1^* & h_0^* & 0 & 0 & 0\\ 0 & h_0 & h_1 & h_2 & 0 & 0\\ 0 & 0 & h_2^* & -h_1^* & h_0^* & 0\\ 0 & 0 & 0 & h_0 & h_1 & h_2\\ 0 & 0 & 0 & 0 & h_2^* & -h_1^*\\ 0 & 0 & 0 & 0 & 0 & h_0\end{bmatrix}.$$

Let the grouping scheme be $\mathcal{I} = \{\mathcal{I}_0 = \{0,2,4\}, \mathcal{I}_1 = \{1,3,5\}\}$. It is easy to verify that

$$\boldsymbol{g}_i \perp \boldsymbol{g}_j,\ i = 0,2,4,\ j = 1,3,5.$$

Similar to Example 2, the system can be decomposed into two systems without performance degrading. For general overlapped Alamouti codes, if we choose the grouping scheme as

$$\begin{cases} \mathcal{I} = \{\mathcal{I}_0 = \{0,2,4,\ldots,n-2\}, \mathcal{I}_1 = \{1,3,5,\ldots,n-1\}\}, \ n \text{ even}, \\ \mathcal{I} = \{\mathcal{I}_0 = \{0,2,4,\ldots,n-1\}, \mathcal{I}_1 = \{1,3,5,\ldots,n-2\}\}, n \text{ odd}. \end{cases}$$

then the system can always be decomposed into two systems without performance degrading. This property is the reason why overlapped Alamouti codes perform better than Toeplitz codes, since the interference comes from only half of the symbols.

## IV. Full Diversity Criterion For PIC and PIC-SIC Group Decodings

In this section, we propose a design criterion for linear dispersion STBC to achieve full diversity with the PIC and the PIC-SIC group decodings.

### *A. Notations and Definitions*

For convenience, let us first introduce some notations and definitions. Let $\mathcal{S}$ be a subset of the complex number field $\mathbb{C}$, we define the difference set $\Delta\mathcal{S}$ as follows,

$$\Delta\mathcal{S} = \{a - \tilde{a}, \ \big| \ a, \tilde{a} \in \mathcal{S}\}.$$

We also introduce the following definition, which can be viewed as an extension of the conventional linear independence concept.

**Definition 1.** *Let $\mathcal{S}$ be a subset of $\mathbb{C}$ and $\boldsymbol{v}_i \in \mathbb{C}^m, i = 0,1,\ldots,n-1$, be $n$ complex vectors. Vectors $\boldsymbol{v}_0, \boldsymbol{v}_1, \ldots, \boldsymbol{v}_{n-1}$ are called* linearly dependent over *$\mathcal{S}$ if there exist $a_0, a_1, \ldots, a_{n-1} \in \mathcal{S}$ so that*

$$a_0\boldsymbol{v}_0 + a_1\boldsymbol{v}_1 + \cdots + a_{n-1}\boldsymbol{v}_{n-1} = \mathbf{0}, \tag{27}$$

*where $a_0, a_1, \ldots, a_{n-1}$ are not all zero; otherwise, vectors $\boldsymbol{v}_0, \boldsymbol{v}_1, \ldots, \boldsymbol{v}_{n-1}$ are called* linear independent over *$\mathcal{S}$.*

For diversity order, the following definition is known.

**Definition 2.** *Consider a communication system as described in* (9). *The system is said to achieve diversity order $m$ if the symbol error rate $\boldsymbol{P}_{\mathsf{SER}}(\mathsf{SNR})$ decays as the inverse of the $m$-th power of* SNR, *i.e.,*

$$\boldsymbol{P}_{\mathsf{SER}}(\mathsf{SNR}) \le c \cdot \mathsf{SNR}^{-m},$$

*where $c > 0$ is a constant independent of* $\mathsf{SNR}$.

The conventional concepts of linear independence and orthogonality are defined among vectors. Next, we define them among vector groups.

**Definition 3.** *Let $\mathcal{V} = \{\boldsymbol{v}_i \in \mathbb{C}^n, i = 0, 1, 2, \dots, k-1\}$ be a set of vectors. Vector $\boldsymbol{v}_k$ is said to be* independent of $\mathcal{V}$ *if for any $a_i \in \mathbb{C}, i = 0, 1, \dots, k-1$,*

$$\boldsymbol{v}_k - \sum_{i=0}^{k-1} a_i \boldsymbol{v}_i \neq \mathbf{0}.$$

*Vector $\boldsymbol{v}_k$ is said to be* orthogonal to $\mathcal{V}$ *if $\boldsymbol{v}_k \perp \boldsymbol{v}_i$, $i = 0, 1, \dots, k-1$.*

**Definition 4.** *Let $\mathcal{V}_0, \mathcal{V}_1, \dots, \mathcal{V}_{n-1}, \mathcal{V}_n$ be $n+1$ groups of vectors. Vector group $\mathcal{V}_n$ is said to be* independent of $\mathcal{V}_0, \mathcal{V}_1, \dots, \mathcal{V}_{n-1}$ *if every vector in $\mathcal{V}_n$ is independent of $\bigcup_{i=0}^{n-1} \mathcal{V}_i$. Vector group $\mathcal{V}_n$ is said to be* orthogonal to $\mathcal{V}_0, \mathcal{V}_1, \dots, \mathcal{V}_{n-1}$ *if every vector in $\mathcal{V}_n$ is orthogonal to $\bigcup_{i=0}^{n-1} \mathcal{V}_i$. Vector groups $\mathcal{V}_0, \mathcal{V}_1, \dots, \mathcal{V}_n$ are said to be linearly independent if for $0 \leq k \leq n$, $\mathcal{V}_k$ is independent of the remaining vector groups $\mathcal{V}_0, \mathcal{V}_1, \dots, \mathcal{V}_{k-1}, \mathcal{V}_{k+1}, \dots, \mathcal{V}_n$. Vector groups $\mathcal{V}_0, \mathcal{V}_1, \dots, \mathcal{V}_n$ are said to be orthogonal if for $0 \leq k \leq n$, $\mathcal{V}_k$ is orthogonal to the remaining vector groups $\mathcal{V}_0, \mathcal{V}_1, \dots, \mathcal{V}_{k-1}, \mathcal{V}_{k+1}, \dots, \mathcal{V}_n$.*

In the remaining of this paper, for convenience, a matrix notation such as $\boldsymbol{G}$ is also used to denote the vector group that is composed of all the column vectors of $\boldsymbol{G}$.

### *B. Design Criterion of STBC with the PIC Group Decoding*

In this subsection, we derive a design criterion of codes decoded with the PIC group decoding. First we introduce the following lemma, which gives a sufficient condition to achieve full diversity for the general channel model in (9) with the ML receiver.

**Lemma 3.** *Consider a communication system modeled as in* (9)*. $\mathcal{A}$ is a signal constellation used in the system. If the channel matrix $\boldsymbol{G}(\boldsymbol{h})$ satisfies the following inequality,*

$$\|\boldsymbol{G}(\boldsymbol{h})\Delta\boldsymbol{x}\|^2 \geq c \cdot \sum_{i=0}^{l-1} |h_i|^2 \, \|\Delta\boldsymbol{x}\|^2, \ \ \forall \Delta\boldsymbol{x} \in \Delta\mathcal{A}^n,$$

*for some positive constant $c$, then the system achieves diversity order $l$ with the ML receiver.*

The proof of this lemma is simply a matter of computation of some integrals, which is quite similar to those derivations in [13], [37]. A detailed proof is given in Appendix C. To understand the meaning of Lemma 3, let us first define the power gain for the channel model in (9).

**Definition 5.** *Consider the communication system modeled as in* (9). $\mathcal{A}$ *is a signal constellation used in the system. The power gain of the system is defined as*

$$P = \min_{\Delta \boldsymbol{x} \in \Delta \mathcal{A}^n} \frac{\|\boldsymbol{G}(\boldsymbol{h})\Delta \boldsymbol{x}\|^2}{\|\Delta \boldsymbol{x}\|^2}.$$

*If the power gain* $P$ *satisfies the following inequality,*

$$P \geq c \cdot \sum_{i=0}^{l-1} |h_i|^2 ,$$

*for some positive constant* $c$*, then we say that the system achieves power gain order* $l$.

From Lemma 3, one can see that the diversity order is ensured by the above power gain order and it can be further interpreted as follows. Suppose that there are two different symbol vectors $\boldsymbol{x}_0, \boldsymbol{x}_1 \in \mathcal{A}^n$. The distance between the two symbol vectors is $\|\Delta \boldsymbol{x}\| \triangleq \|\boldsymbol{x}_0 - \boldsymbol{x}_1\|$. Assume there is no noise in the channel, i.e., $\boldsymbol{w} = \boldsymbol{0}$, then after the symbol vectors pass through the channel, we get $\boldsymbol{G}(\boldsymbol{h})\boldsymbol{x}_0$ and $\boldsymbol{G}(\boldsymbol{h})\boldsymbol{x}_1$. Now the distance between received signals $\boldsymbol{G}(\boldsymbol{h})\boldsymbol{x}_0$ and $\boldsymbol{G}(\boldsymbol{h})\boldsymbol{x}_1$ is $\|\boldsymbol{G}(\boldsymbol{h})\Delta \boldsymbol{x}\|$, which is greater than $\sqrt{P}\,\|\Delta \boldsymbol{x}\|$, i.e., the channel "expanded" the distance between $\boldsymbol{x}_0$ and $\boldsymbol{x}_1$ by a factor of at least $\sqrt{P}$. The expansion factor $\sqrt{P}$ determines the diversity order that can be achieved. Lemma 3 tells us that if the expansion factor $\sqrt{P}$ of the symbol vector is greater than $\left(c \cdot \sum_{i=0}^{l-1} |h_i|^2\right)^{\frac{1}{2}}$ for some $c > 0$, then diversity order $l$ can be achieved. Note that the power gain order can be viewed as a count of how many path gains summed up in $P$. We can rephrase Lemma 3 simply as: if the power gain is a sum of $l$ path gains, then the diversity order of the communication system in (9) is $l$.

Next, we present the main result of this paper, which characterizes the power gain order of a linear dispersion STBC decoded with the PIC and the PIC-SIC group decoding algorithms.

**Theorem 1** (Main Theorem)**.** *Let* $\mathcal{X}$ *be a linear dispersion STBC. There are* $n_t$ *transmit and* $n_r$ *receive antennas. The channel matrix is* $\boldsymbol{H} \in \mathbb{C}^{n_r \times n_t}$. *The received signal is decoded using the PIC group decoding with a grouping scheme* $\{\mathcal{I}_0, \mathcal{I}_1, \ldots, \mathcal{I}_{N-1}\}$. *The equivalent channel is* $\boldsymbol{G}(\boldsymbol{h})$, *where* $\boldsymbol{h} = \mathrm{vec}(\boldsymbol{H}) = \{h_0, h_1, \ldots, h_{n_r \cdot n_t - 1}\} \in \mathbb{C}^{n_r \cdot n_t}$. *Then, each of the following* *dimension-reduced systems (i.e., the STBC with the PIC group decoding),*

$$\boldsymbol{z}_{\mathcal{I}_k} = \boldsymbol{P}_{\mathcal{I}_k}\boldsymbol{G}_{\mathcal{I}_k}\boldsymbol{x}_{\mathcal{I}_k} + \boldsymbol{P}_{\mathcal{I}_k}\boldsymbol{w},\ k = 0, 1, \ldots, N-1, \tag{28}$$

*has power gain order* $n_r \cdot n_t$ *if and only if the following two conditions are satisfied:*

- *for any two different codewords* $\boldsymbol{X}, \tilde{\boldsymbol{X}} \in \mathcal{X}$, $\Delta\boldsymbol{X} \triangleq \boldsymbol{X} - \tilde{\boldsymbol{X}}$ *has the full rank property, i.e., the code* $\mathcal{X}$ *achieves full diversity with the ML receiver;*
- $\boldsymbol{G}_{\mathcal{I}_0}, \boldsymbol{G}_{\mathcal{I}_1}, \ldots, \boldsymbol{G}_{\mathcal{I}_{N-1}}$ *defined in* (10) *from* $\boldsymbol{G} = \boldsymbol{G}(\boldsymbol{h})$ *are linearly independent vector groups as long as* $\boldsymbol{h} \neq \boldsymbol{0}$.

*When the received signals are decoded using the PIC-SIC group decoding with the ordering (24), each dimension-reduced system derived during the decoding process (i.e., the STBC with the PIC-SIC group decoding) has power gain order* $n_r \cdot n_t$ *if and only if*

- *for any two different codewords* $\boldsymbol{X}, \tilde{\boldsymbol{X}} \in \mathcal{X}$, $\Delta\boldsymbol{X} \triangleq \boldsymbol{X} - \tilde{\boldsymbol{X}}$ *has the full rank property, i.e., the code* $\mathcal{X}$ *achieves full diversity with the ML receiver;*
- *at each decoding stage,* $\boldsymbol{G}_{\mathcal{I}_{i_k}}$*, which corresponds to the current to-be decoded symbol group* $\boldsymbol{x}_{i_k}$*, and* $[\boldsymbol{G}_{\mathcal{I}_{i_{k+1}}}, \ldots, \boldsymbol{G}_{\mathcal{I}_{i_{N-1}}}]$ *are linearly independent vector groups as long as* $\boldsymbol{h} \neq \boldsymbol{0}$.

With the above theorem and the preceding discussions on the relationship between diversity order and power gain order, the two conditions in the above theorem provide a criterion for a linear dispersion code to achieve full diversity with the PIC group decoding.

Let us see an example to use the above main theorem. Consider the rotated quasi-orthogonal scheme proposed in [35] for a QAM signal constellation, where the code $\boldsymbol{X}$ has the following structure,

$$\boldsymbol{X} = \begin{bmatrix} x_0 & -x_1^* & \alpha x_2 & -\alpha^* x_3^* \\ x_1 & x_0^* & \alpha x_3 & \alpha^* x_2^* \\ \alpha x_2 & -\alpha^* x_3^* & x_0 & -x_1^* \\ \alpha x_3 & \alpha^* x_2^* & x_1 & x_0^* \end{bmatrix},\ \alpha = \exp\left(\frac{\boldsymbol{i}\pi}{4}\right). \tag{29}$$

Suppose we use one receive antenna, the column vectors of the equivalent channel $\boldsymbol{G}$ are as follows,

$$\boldsymbol{g}_0 = \frac{1}{2}\begin{bmatrix} h_0 \\ h_1^* \\ h_2 \\ h_3^* \end{bmatrix},\ \boldsymbol{g}_1 = \frac{1}{2}\begin{bmatrix} h_1 \\ -h_0^* \\ h_3 \\ -h_2^* \end{bmatrix},\ \boldsymbol{g}_2 = \frac{1}{2}\begin{bmatrix} \alpha h_2 \\ \alpha h_3^* \\ \alpha h_0 \\ \alpha h_1^* \end{bmatrix},\ \boldsymbol{g}_3 = \frac{1}{2}\begin{bmatrix} \alpha h_3 \\ -\alpha h_2^* \\ \alpha h_1 \\ -\alpha h_0^* \end{bmatrix}. \tag{30}$$

It has also been proved in [35] that this code achieves full diversity with the ML receiver, hence the first condition is satisfied. Let the grouping scheme be $\{\mathcal{I}_0 = \{0, 2\}, \mathcal{I}_1 = \{1, 3\}\}$, then $\boldsymbol{G}_{\mathcal{I}_0}$ and $\boldsymbol{G}_{\mathcal{I}_1}$ are linearly independent. Thus, both conditions are satisfied. Note that the two groups are actually orthogonal, which means that every vector in $\boldsymbol{G}_{\mathcal{I}_0}$ is orthogonal to $\boldsymbol{G}_{\mathcal{I}_1}$ and vice versa. Hence after the interference cancellation, there is no power gain loss. In this case, the PIC group decoding is exactly the same as the ML receiver.

### *C. Proof of the Main Theorem*

In order to prove the main theorem, let us first introduce the following lemma.

**Lemma 4.** *Consider a communication system modeled as in* (9). *$\mathcal{A}$ is a signal constellation used in the system. If the equivalent channel matrix $\boldsymbol{G}(\boldsymbol{h})$ satisfies the following two conditions:*

- *scaling invariance:*

$$\frac{1}{\|\boldsymbol{h}\|}\boldsymbol{G}(\boldsymbol{h}) = \boldsymbol{G}\left(\frac{\boldsymbol{h}}{\|\boldsymbol{h}\|}\right); \tag{31}$$

- *the column vectors of $\boldsymbol{G}(\boldsymbol{h})$ are linearly independent over $\Delta\mathcal{A}$ for any $\mathbf{0} \neq \boldsymbol{h} \in \mathbb{C}^l$,*

*then the system has power gain order $l$ and thus achieves diversity order $l$ with the ML receiver.*

A proof is given in Appendix D. Note that if each entry of $\boldsymbol{G}(\boldsymbol{h})$ is a linear combination of $h_0, h_1, \ldots, h_{l-1}$ and $h_0^*, h_1^*, \ldots, h_{l-1}^*$, then the scaling invariance (31) always holds. So we have the following corollary.

**Corollary 1.** *Consider a communication system modeled as in* (9). *Each entry of $\boldsymbol{G}(\boldsymbol{h})$ is a linear combination of $h_0, h_1, \ldots, h_{l-1}$ and $h_0^*, h_1^*, \ldots, h_{l-1}^*$. $\mathcal{A}$ is a signal constellation used in the system. If the column vectors of $\boldsymbol{G}(\boldsymbol{h})$ are linearly independent over $\Delta\mathcal{A}$ for any $\mathbf{0} \neq \boldsymbol{h} = [h_0, h_1, \ldots, h_{l-1}]^{\mathsf{T}} \in \mathbb{C}^l$, then the system has power gain order $l$ thus achieves diversity order $l$ with the ML receiver.*

One may wonder for linear dispersion STBC, whether the above condition is an equivalent condition of the *full rank* criterion. The following theorem gives a positive answer to this question.

**Theorem 2.** *Let $\mathcal{X}$ be a linear dispersion STBC. Let $\mathcal{A}$ be a signal constellation for the coding scheme $\mathcal{X}$. Let $\boldsymbol{G}(\boldsymbol{h})$ be the equivalent channel of $\mathcal{X}$ and $\boldsymbol{h}$ and $\boldsymbol{h} \neq \mathbf{0}$. Then $\mathcal{X}$ has the* full rank *property if and only if the column vectors of $\boldsymbol{G}(\boldsymbol{h})$ are linearly independent over $\Delta\mathcal{A}$.*

Its proof is in Appendix E.

Now we are ready to prove the main theorem. The main idea is to prove that the dimension-reduced systems in (28) satisfy the two conditions in Lemma 4.

*1) Sufficiency part:* First we prove that the two conditions in the main theorem are sufficient conditions for codes to achieve the full power gain with the PIC group decoding algorithm. According to Theorem 2, the first condition is equivalent to that the column vectors of $\boldsymbol{G}(\boldsymbol{h})$ are linearly independent over $\Delta\mathcal{A}$. This further implies that the column vectors of $\boldsymbol{G}_{\mathcal{I}_k}$ are linearly independent over $\Delta\mathcal{A}$, i.e., for any $a_0, a_1, \ldots, a_{n_k-1} \in \Delta\mathcal{A}$, $a_j, j = 0, 1, \ldots, n_k - 1$, not all zero, we have

$$\sum_{j=0}^{n_k-1} a_j \boldsymbol{g}_{i_{k,j}} \neq \boldsymbol{0}. \tag{32}$$

Since $\boldsymbol{G}_{\mathcal{I}_0}, \boldsymbol{G}_{\mathcal{I}_1}, \ldots, \boldsymbol{G}_{\mathcal{I}_{N-1}}$ are linearly independent, the column vectors $\boldsymbol{g}_{i_{k,j}}, j = 0, 1, \ldots, n_k-1$, in $\boldsymbol{G}_{\mathcal{I}_k}$ do not belong[1] to the vector space $\mathbb{V}_{\mathcal{I}_k}$ defined in (12). From (32) and the fact that $\boldsymbol{g}_{i_{k,j}} \notin \mathbb{V}_{\mathcal{I}_k}$, we have

$$\boldsymbol{Q}_{\mathcal{I}_k}\left(\sum_{j=0}^{n_k-1} a_j \boldsymbol{g}_{i_{k,j}}\right) \neq \sum_{j=0}^{n_k-1} a_j \boldsymbol{g}_{i_{k,j}}.$$

By applying the above inequality, we get the following inequality,

$$\begin{aligned}\sum_{j=0}^{n_k-1} a_j \boldsymbol{P}_{\mathcal{I}_k} \boldsymbol{g}_{i_{k,j}} &= \boldsymbol{P}_{\mathcal{I}_k}\left(\sum_{j=0}^{n_k-1} a_j \boldsymbol{g}_{i_{k,j}}\right) \\ &= (\boldsymbol{I}_m - \boldsymbol{Q}_{\mathcal{I}_k}) \sum_{j=0}^{n_k-1} a_j \boldsymbol{g}_{i_{k,j}} \neq \boldsymbol{0},\end{aligned}$$

i.e., the column vectors of $\boldsymbol{P}_{\mathcal{I}_k}\boldsymbol{G}_{\mathcal{I}_k}$ are also linearly independent over $\Delta\mathcal{A}$.

Now we prove that $\boldsymbol{P}_{\mathcal{I}_k}\boldsymbol{G}_{\mathcal{I}_k}$ satisfies the scaling invariance (31) in Lemma 4. Since both $\boldsymbol{P}_{\mathcal{I}_k}$ and $\boldsymbol{G}_{\mathcal{I}_k}$ are determined by the parameter vector $\boldsymbol{h}$, for a clear exposition, we temporarily use $\boldsymbol{P}_{\mathcal{I}_k}(\boldsymbol{h})$ to denote $\boldsymbol{P}_{\mathcal{I}_k}$ and use $\boldsymbol{G}_{\mathcal{I}_k}(\boldsymbol{h})$ to denote $\boldsymbol{G}_{\mathcal{I}_k}$. Then we have

$$\begin{aligned}\frac{1}{\|\boldsymbol{h}\|}\boldsymbol{P}_{\mathcal{I}_k}(\boldsymbol{h})\boldsymbol{G}_{\mathcal{I}_k}(\boldsymbol{h}) &= \boldsymbol{P}_{\mathcal{I}_k}(\boldsymbol{h})\left(\frac{1}{\|\boldsymbol{h}\|}\boldsymbol{G}_{\mathcal{I}_k}(\boldsymbol{h})\right) \\ &= \boldsymbol{P}_{\mathcal{I}_k}(\boldsymbol{h})\boldsymbol{G}_{\mathcal{I}_k}\left(\frac{\boldsymbol{h}}{\|\boldsymbol{h}\|}\right) \\ &= \boldsymbol{P}_{\mathcal{I}_k}\left(\frac{\boldsymbol{h}}{\|\boldsymbol{h}\|}\right)\boldsymbol{G}_{\mathcal{I}_k}\left(\frac{\boldsymbol{h}}{\|\boldsymbol{h}\|}\right),\end{aligned}$$

[1] Here the linear independence over the whole complex field of the vector sets is needed/used and the linear independence over $\Delta\mathcal{A}$ is not sufficient.

where the second equality holds since the entries in $\boldsymbol{G}_{\mathcal{I}_k}$ are all linear combinations of $h_0, h_1, \ldots, h_{n_r \cdot n_t}$ and $h_0^*, h_1^*, \ldots, h_{n_r \cdot n_t}^*$, and the last equality holds since

$$\boldsymbol{P}_{\mathcal{I}_k}(\boldsymbol{h}) = \boldsymbol{P}_{\mathcal{I}_k}\left(\frac{\boldsymbol{h}}{\|\boldsymbol{h}\|}\right),$$

which is a direct result from the definition of $\boldsymbol{Q}_{\mathcal{I}_k}$ in (14) and the fact that $\boldsymbol{P}_{\mathcal{I}_k} = \boldsymbol{I}_m - \boldsymbol{Q}_{\mathcal{I}_k}$.

Thus, the two conditions in Lemma 4 are all satisfied and therefore for any $k$, the dimension-reduced system

$$\boldsymbol{z}_{\mathcal{I}_k} = \sqrt{\mathsf{SNR}}\left(\boldsymbol{P}_{\mathcal{I}_k}\boldsymbol{G}_{\mathcal{I}_k}\right)\boldsymbol{x}_{\mathcal{I}_k} + \boldsymbol{P}_{\mathcal{I}_k}\boldsymbol{w},$$

has power gain order $n_r \cdot n_t$.

Now let us consider the case when the received signals are decoded with the PIC-SIC group decoding. We use the conventional assumption that the previous decoded symbols are correct. Thus, there is no error introduced when we use these decoded symbols to reduce the interferences from the received signals. Under this assumption, the PIC-SIC group decoding algorithm is always better than the PIC group decoding algorithm. Thus, the two conditions are sufficient for the PIC-SIC case.

*2) Necessity part:* We next prove that these two conditions are also necessary conditions. If $\boldsymbol{G}_{\mathcal{I}_k}$ and $\boldsymbol{G}_{\mathcal{I}_k}^c = [\boldsymbol{G}_{\mathcal{I}_0}, \ldots, \boldsymbol{G}_{\mathcal{I}_{k-1}}, \boldsymbol{G}_{\mathcal{I}_{k+1}}, \ldots, \boldsymbol{G}_{\mathcal{I}_{N-1}}]$ are not linearly independent, i.e., there exists a column vector in $\boldsymbol{G}_{\mathcal{I}_k}$ such that this vector belongs to the subspace $\mathbb{V}_{\mathcal{I}_k}$. Without loss of generality, we assume this vector is $\boldsymbol{g}_{i_{k,0}}$. In this case, we have

$$\boldsymbol{P}_{\mathcal{I}_k}\boldsymbol{G}_{\mathcal{I}_k} = \left[\boldsymbol{0}, \boldsymbol{P}_{\mathcal{I}_k}\boldsymbol{g}_{i_{k,1}}, \boldsymbol{P}_{\mathcal{I}_k}\boldsymbol{g}_{i_{k,2}}, \ldots, \boldsymbol{P}_{\mathcal{I}_k}\boldsymbol{g}_{i_{k,n_k-1}}\right].$$

Take $\Delta\boldsymbol{x}_{\mathcal{I}_k} = [a, 0, 0, \ldots, 0]^{\mathsf{T}} \in \Delta\mathcal{A}^{n_k}$, where $a \in \Delta\mathcal{A}, a \neq 0$, then we have $\|\boldsymbol{P}_{\mathcal{I}_k}\boldsymbol{G}_{\mathcal{I}_k}\Delta\boldsymbol{x}_{\mathcal{I}_k}\| = \boldsymbol{0}$, which contradicts with the condition that the systems in (28) have power gain order $n_r \cdot n_t$. Thus, we must have that $\boldsymbol{G}_{\mathcal{I}_k}$ is linearly independent of $\boldsymbol{G}_{\mathcal{I}_k}^c$. Since $k$ is an arbitrary integer number in $[0, N-1]$, $\boldsymbol{G}_{\mathcal{I}_0}, \boldsymbol{G}_{\mathcal{I}_1}, \ldots, \boldsymbol{G}_{\mathcal{I}_{N-1}}$ are linearly independent. This proves that the second condition in the main theorem must hold.

Let $\Delta\boldsymbol{x} \neq \boldsymbol{0} \in \Delta\mathcal{A}^n$ and $\Delta\boldsymbol{x}_{\mathcal{I}_k}, k = 0, 1, \ldots, N-1$, be the corresponding sub-vectors of $\Delta\boldsymbol{x}$ to the grouping scheme. Thus, there is at least one $\Delta\boldsymbol{x}_{\mathcal{I}_k} \neq \boldsymbol{0}$. Without loss of generality, we assume $\Delta \boldsymbol{x}_{\mathcal{I}_0} \neq \mathbf{0}$. Then,

$$\begin{aligned}\|\boldsymbol{G}(\boldsymbol{h})\Delta\boldsymbol{x}\|^2 &= \left\|\boldsymbol{G}_{\mathcal{I}_0}\Delta\boldsymbol{x}_{\mathcal{I}_0} + \sum_{i=1}^{N-1}\boldsymbol{G}_{\mathcal{I}_k}\Delta\boldsymbol{x}_{\mathcal{I}_k}\right\|^2 \\ &= \left\|\boldsymbol{P}_{\mathcal{I}_0}\boldsymbol{G}_{\mathcal{I}_0}\Delta\boldsymbol{x}_{\mathcal{I}_0} + \boldsymbol{Q}_{\mathcal{I}_0}\boldsymbol{G}_{\mathcal{I}_0}\Delta\boldsymbol{x}_{\mathcal{I}_0} + \sum_{i=1}^{N-1}\boldsymbol{G}_{\mathcal{I}_k}\Delta\boldsymbol{x}_{\mathcal{I}_k}\right\|^2.\end{aligned}$$

Since $\boldsymbol{P}_{\mathcal{I}_0}\boldsymbol{G}_{\mathcal{I}_0}\Delta\boldsymbol{x}_{\mathcal{I}_0} \in \mathbb{V}_{\mathcal{I}_0}^{\perp}$ and $\boldsymbol{Q}_{\mathcal{I}_0}\boldsymbol{G}_{\mathcal{I}_0}\Delta\boldsymbol{x}_{\mathcal{I}_0} + \sum_{i=1}^{N-1}\boldsymbol{G}_{\mathcal{I}_k}\Delta\boldsymbol{x}_{\mathcal{I}_k} \in \mathbb{V}_{\mathcal{I}_0}$, we have

$$\begin{aligned}\|\boldsymbol{G}(\boldsymbol{h})\Delta\boldsymbol{x}\|^2 &= \|\boldsymbol{P}_{\mathcal{I}_0}\boldsymbol{G}_{\mathcal{I}_0}\Delta\boldsymbol{x}_{\mathcal{I}_0}\|^2 + \left\|\boldsymbol{Q}_{\mathcal{I}_0}\boldsymbol{G}_{\mathcal{I}_0}\Delta\boldsymbol{x}_{\mathcal{I}_0} + \sum_{i=1}^{N-1}\boldsymbol{G}_{\mathcal{I}_k}\Delta\boldsymbol{x}_{\mathcal{I}_k}\right\|^2 \\ &\geq \|\boldsymbol{P}_{\mathcal{I}_0}\boldsymbol{G}_{\mathcal{I}_0}\Delta\boldsymbol{x}_{\mathcal{I}_0}\|^2 \\ &\geq c\cdot\left(\sum_{i=0}^{n_r\cdot n_t-1}|h_i|^2\right)\|\Delta\boldsymbol{x}_{\mathcal{I}_0}\|^2 > 0,\ \boldsymbol{h}\neq\mathbf{0}.\end{aligned}$$

Using Theorem 2, the first condition in the theorem is proved.

In the case that the received signals are decoded with the PIC-SIC group decoding, we assume the decoding order is $\mathcal{I}_{i_0}, \mathcal{I}_{i_1}, \ldots, \mathcal{I}_{i_{N-1}}$. Similar to the above argument, we must have that $\boldsymbol{G}_{\mathcal{I}_{i_0}}$ is linearly independent of $\boldsymbol{G}_{\mathcal{I}_{i_1}}, \boldsymbol{G}_{\mathcal{I}_{i_2}}, \ldots, \boldsymbol{G}_{\mathcal{I}_{i_{N-1}}}$; $\boldsymbol{G}_{\mathcal{I}_{i_1}}$ is linearly independent of $\boldsymbol{G}_{\mathcal{I}_{i_2}}, \boldsymbol{G}_{\mathcal{I}_{i_3}}, \ldots, \boldsymbol{G}_{\mathcal{I}_{i_{N-1}}}$; $\boldsymbol{G}_{\mathcal{I}_{i_2}}$ is linearly independent of $\boldsymbol{G}_{\mathcal{I}_{i_3}}, \boldsymbol{G}_{\mathcal{I}_{i_4}}, \ldots, \boldsymbol{G}_{\mathcal{I}_{i_{N-1}}}$ etc. So we have that $\boldsymbol{G}_{\mathcal{I}_{i_0}}, \boldsymbol{G}_{\mathcal{I}_{i_1}}, \ldots, \boldsymbol{G}_{\mathcal{I}_{i_{N-1}}}$ are linearly independent. The proof of the first condition to be necessary is the same as the PIC case. This completes our proof of the main theorem.

### *D. Connection with the Full Rank Criterion and the Shang-Xia Criterion*

In the case when there is only one group, then the PIC group decoding algorithm becomes the ML decoding. In this case the second condition can always be satisfied. Thus, our proposed design criterion in Theorem 1 is equivalent to that of [13], [37].

We now consider the symbol-by-symbol grouping case of the PIC group decoding algorithm, which is equivalent to the ZF decoding algorithm. In this case when each group contains only one symbol, the second condition can be rephrased as: $\boldsymbol{G}(\boldsymbol{h})$ is a column full rank matrix for $\boldsymbol{h} \neq \mathbf{0}$.

**Corollary 2.** *In the case of symbol-by-symbol PIC group decoding, i.e., each group only contains one symbol, the design criterion in the main theorem is equivalent to the Shang-Xia criterion* *proposed in [32], i.e.,*

$$\det\left(\boldsymbol{G}(\boldsymbol{h})^{\mathsf{H}}\boldsymbol{G}(\boldsymbol{h})\right) \geq c\left\|\boldsymbol{h}\right\|^{2n},\ \boldsymbol{h} \in \mathbb{C}^l,$$

*where $c$ is a constant independent of the channel $\boldsymbol{h}$.*

*Proof:* Since we have that $\boldsymbol{G}(\boldsymbol{h})$ is full column rank for $\boldsymbol{h} \neq \boldsymbol{0}$, the following inequality must hold,

$$\det\left(\boldsymbol{G}(\boldsymbol{h})^{\mathsf{H}}\boldsymbol{G}(\boldsymbol{h})\right) > 0,\ \boldsymbol{h} \neq \boldsymbol{0},$$

Let us restrict the parameter $\boldsymbol{h}$ to the unit sphere, i.e., $\|\boldsymbol{h}\| = 1$. Note that the unit sphere is a compact set, $\det\left(\boldsymbol{G}(\boldsymbol{h})^{\mathsf{H}}\boldsymbol{G}(\boldsymbol{h})\right)$ is a continuous function of $\boldsymbol{h}$. There must exist a positive constant $c > 0$ such at

$$\det\left(\boldsymbol{G}(\boldsymbol{h})^{\mathsf{H}}\boldsymbol{G}(\boldsymbol{h})\right) > c,\ \boldsymbol{h} \neq \boldsymbol{0},$$

as what is used in [48]. Generally, for $\boldsymbol{h} \in \mathbb{C}^l \setminus \{\boldsymbol{0}\}$, we have that

$$\det\left(\boldsymbol{G}\left(\frac{\boldsymbol{h}}{\|\boldsymbol{h}\|}\right)^{\mathsf{H}}\boldsymbol{G}\left(\frac{\boldsymbol{h}}{\|\boldsymbol{h}\|}\right)\right) > c,\ \boldsymbol{h} \neq \boldsymbol{0}. \tag{33}$$

Since the entries of $\boldsymbol{G}(\boldsymbol{h})$ are linear combinations of $h_0, h_1, \ldots, h_{l-1}$ and $h_0^*, h_1^*, \ldots, h_{l-1}^*$, inequality (33) can be rewritten as

$$\det\left(\frac{\boldsymbol{G}(\boldsymbol{h})^{\mathsf{H}}\boldsymbol{G}(\boldsymbol{h})}{\|\boldsymbol{h}\|^2}\right) > c,\ \boldsymbol{h} \neq \boldsymbol{0}. \tag{34}$$

Thus,

$$\det\left(\boldsymbol{G}(\boldsymbol{h})^{\mathsf{H}}\boldsymbol{G}(\boldsymbol{h})\right) \geq c\left\|\boldsymbol{h}\right\|^{2n},\ \boldsymbol{h} \in \mathbb{C}^l, \tag{35}$$

which is the Shang-Xia condition given in [32]. This proves that the criterion in Theorem 1 implies the Shang-Xia criterion in the case when all symbols are in separate groups, i.e., the ZF receiver.

Since the criterion in Theorem 1 is necessary and sufficient, it can be derived from the Shang-Xia criterion too. In other words, the criterion in Theorem 1 is equivalent to the Shang-Xia criterion in the case when the ZF receiver is used. ∎

### E. Some Discussions

From Theorem 1 and Theorem 2, it is interesting to see that for a linear dispersion STBC (complex conjugates of symbols may be embedded) to achieve full diversity: (i) the weakest criterion is that the column vectors of the equivalent channel matrix are linearly independent over a signal constellation $\mathcal{A}$ when the ML receiver is used, which is equivalent to the code full rank criterion known in the literature; (ii) the strongest criterion (in the sense of the simplest complex-symbol-wise decoding) is that the column vectors of the equivalent channel matrix are linearly independent over the whole complex field when the ZF receiver is used, which is, in fact, weaker than the orthogonality in the OSTBC case that is not necessary for achieving full diversity with a linear receiver. In the case of the weakest criterion but the optimal and the most complicated receiver, i.e., ML receiver, the symbol rate can be $n_t$ for $n_t$ transmit antennas. In the case of the strongest criterion but the simplest receiver, i.e., linear receiver, the symbol rate can not be above 1 [32]. Note that the rates of OSTBC approaches to $1/2$ as the number of transmit antennas goes to infinity and are upper bounded by $3/4$ for more than $2$ transmit antennas [44]. By increasing the decoding complexity and improving a receiver as increasing the group sizes in our proposed PIC group decoding, the criterion to achieve full diversity becomes weaker. The criterion for the PIC group decoding serves as a bridge between the strongest and the weakest criteria for the ZF and the ML receivers, respectively, and the corresponding symbol rates are expected between $1$ and $n_t$. The examples to be presented later in Section VI are some simple examples to show this rate-complexity tradeoff.

## V. Asymptotic Optimal Group Decoding

In this section, we propose an *asymptotic optimal* (AO) group decoding algorithm, which can be viewed as an intermediate decoding between the ML decoding and the MMSE decoding algorithms [3], [41].

### A. Asymptotic Optimal Group Decoding Algorithm

Consider the channel model in (9). Suppose the signals are decoded using a group decoding algorithm, and the grouping scheme is $\mathcal{I} = \{\mathcal{I}_0, \mathcal{I}_1, \mathcal{I}_2, \ldots, \mathcal{I}_{N-1}\}$. Assume the symbols are taken from a signal constellation $\mathcal{A}$ according to the uniform distribution. The optimal way to decode $\boldsymbol{x}_{\mathcal{I}_k}$ from the received signals is to find $\hat{\boldsymbol{x}}_{\mathcal{I}_k} \in \mathcal{A}^{n_k}$ such that

$$\hat{\boldsymbol{x}}_{\mathcal{I}_k} = \arg \max_{\bar{\boldsymbol{x}}_{\mathcal{I}_k} \in \mathcal{A}^{n_k}} P\left(\boldsymbol{y} \mid \bar{\boldsymbol{x}}_{\mathcal{I}_k}\right).$$

To derive the decoding rule, let us first write (11) in the following form,

$$\boldsymbol{y} = \sqrt{\mathsf{SNR}}\boldsymbol{G}_{\mathcal{I}_k}\boldsymbol{x}_{\mathcal{I}_k} + \sqrt{\mathsf{SNR}} \sum_{\substack{0 \le i \le N-1 \\ i \ne k}} \boldsymbol{G}_{\mathcal{I}_i}\boldsymbol{x}_{\mathcal{I}_i} + \boldsymbol{w}. \tag{36}$$

Note that except for the symbol group $\boldsymbol{x}_{\mathcal{I}_k}$, all the other symbols can be viewed as noises that interfere with $\boldsymbol{x}_{\mathcal{I}_k}$. Define the noise term $\boldsymbol{w}_{\mathcal{I}_k}$ as

$$\begin{aligned} \boldsymbol{w}_{\mathcal{I}_k} &= \sqrt{\mathsf{SNR}} \sum_{\substack{0 \le i \le N-1 \\ i \ne k}} \boldsymbol{G}_{\mathcal{I}_i}\boldsymbol{x}_{\mathcal{I}_i} + \boldsymbol{w} \\ &= \sqrt{\mathsf{SNR}} \sum_{i \notin \mathcal{I}_k} \boldsymbol{g}_i x_i + \boldsymbol{w}. \end{aligned} \tag{37}$$

Then, we can write (36) as

$$\boldsymbol{y} = \sqrt{\mathsf{SNR}}\boldsymbol{G}_{\mathcal{I}_k}\boldsymbol{x}_{\mathcal{I}_k} + \boldsymbol{w}_{\mathcal{I}_k}. \tag{38}$$

The optimal decoding of $\boldsymbol{x}_{\mathcal{I}_k}$ from the received signal vector $\boldsymbol{y}$ depends on the distribution of the noise $\boldsymbol{w}_{\mathcal{I}_k}$, which is difficult to analyze in general. To simplify the discussion, we assume that the noise $\boldsymbol{w}_{\mathcal{I}_k}$ is Gaussian. This assumption is asymptotically true when the number of the interference symbols is large. Similar assumption has been used in [24]. We call the optimal result derived under this assumption asymptotically optimal.

Under the above assumption, the probability density function $P\left(\boldsymbol{y} \mid \bar{\boldsymbol{x}}_{\mathcal{I}_k}\right)$ can be explicitly expressed and the optimal decoding rule can be easily derived. First let us compute the covariance matrix of the noise vector $\boldsymbol{w}_{\mathcal{I}_k}$:

$$\begin{aligned} \boldsymbol{K}_{\mathcal{I}_k} &= \mathcal{E}\left\{\boldsymbol{w}_{\mathcal{I}_k}\boldsymbol{w}_{\mathcal{I}_k}^{\mathsf{H}}\right\} \\ &= \boldsymbol{I}_m + \mathsf{SNR} \sum_{i \notin \mathcal{I}_k} \boldsymbol{g}_i \boldsymbol{g}_i^{\mathsf{H}}. \end{aligned}$$

Hence the probability density function $P\left(\boldsymbol{y} \mid \bar{\boldsymbol{x}}_{\mathcal{I}_k}\right)$ is as follows,

$$P\left(\boldsymbol{y} \mid \bar{\boldsymbol{x}}_{\mathcal{I}_k}\right) = \frac{1}{\pi^m \left|\boldsymbol{K}_{\mathcal{I}_k}\right|} \exp\left(-\left(\boldsymbol{y} - \sqrt{\mathsf{SNR}}\boldsymbol{G}_{\mathcal{I}_k}\boldsymbol{x}_{\mathcal{I}_k}\right)^{\mathsf{H}} \boldsymbol{K}_{\mathcal{I}_k}^{-1} \left(\boldsymbol{y} - \sqrt{\mathsf{SNR}}\boldsymbol{G}_{\mathcal{I}_k}\boldsymbol{x}_{\mathcal{I}_k}\right)\right).$$

For the above equation, we can see that maximizing $P\left(\boldsymbol{y} \mid \bar{\boldsymbol{x}}_{\mathcal{I}_k}\right)$ is equivalent to minimizing

$$\left(\boldsymbol{y} - \sqrt{\mathsf{SNR}}\boldsymbol{G}_{\mathcal{I}_k}\boldsymbol{x}_{\mathcal{I}_k}\right)^{\mathsf{H}} \boldsymbol{K}_{\mathcal{I}_k}^{-1} \left(\boldsymbol{y} - \sqrt{\mathsf{SNR}}\boldsymbol{G}_{\mathcal{I}_k}\boldsymbol{x}_{\mathcal{I}_k}\right) = \left\|\boldsymbol{K}_{\mathcal{I}_k}^{-\frac{1}{2}} \left(\boldsymbol{y} - \sqrt{\mathsf{SNR}}\boldsymbol{G}_{\mathcal{I}_k}\boldsymbol{x}_{\mathcal{I}_k}\right)\right\|^2,$$

where $\boldsymbol{K}_{\mathcal{I}_k}^{-\frac{1}{2}}$ is the square root of the matrix $\boldsymbol{K}_{\mathcal{I}_k}^{-1}$. So the asymptotic optimal decoding rule is

$$\hat{\boldsymbol{x}}_{\mathcal{I}_k} = \arg\max_{\bar{\boldsymbol{x}}_{\mathcal{I}_k} \in \mathcal{A}^{n_k}} \left\| \boldsymbol{K}_{\mathcal{I}_k}^{-\frac{1}{2}} \left( \boldsymbol{y} - \sqrt{\mathsf{SNR}} \boldsymbol{G}_{\mathcal{I}_k} \bar{\boldsymbol{x}}_{\mathcal{I}_k} \right) \right\|. \tag{39}$$

When $\mathcal{I} = \{\mathcal{I}_0\} = \{\{0, 1, 2, \ldots, n-1\}\}$, we only have one symbol group $\boldsymbol{x}_{\mathcal{I}_0}$, which contains all the symbols. The variance of the noise is $\boldsymbol{K}_{\mathcal{I}_0} = I_m$. In this case, the above decoding rule can be simplified as

$$\hat{\boldsymbol{x}} = \hat{\boldsymbol{x}}_{\mathcal{I}_0} = \arg\max_{\bar{\boldsymbol{x}} \in \mathcal{A}^n} \left\| \boldsymbol{y} - \sqrt{\mathsf{SNR}} \boldsymbol{G}(\boldsymbol{h}) \bar{\boldsymbol{x}} \right\|,$$

which is the ML decoding.

### *B. Connection with the MMSE Decoding*

Now let us consider the symbol-by-symbol case of the AO group decoding algorithm. In this case, $\mathcal{I} = \mathcal{I}_0, \mathcal{I}_1, \ldots, \mathcal{I}_{n-1} = \{\{0\}, \{1\}, \ldots, \{n-1\}\}$. In the following discussion, we use the simplified notation convention introduced in III-B. Thus, we use $\boldsymbol{K}_k$ instead of $\boldsymbol{K}_{\mathcal{I}_k}$ to denote

$$\boldsymbol{K}_k = \boldsymbol{I}_m + \mathsf{SNR} \sum_{i \neq k} \boldsymbol{g}_i \boldsymbol{g}_i^{\mathsf{H}}.$$

So the decoding rule is

$$\begin{aligned} \hat{x}_{\mathcal{I}_k} &= \arg\max_{\bar{x}_k \in \mathcal{A}} \left\| \boldsymbol{K}_k^{-\frac{1}{2}} \left( \boldsymbol{y} - \sqrt{\mathsf{SNR}} \boldsymbol{g}_k \bar{x}_k \right) \right\| \\ &= \arg\max_{\bar{x}_k \in \mathcal{A}} \left\| \frac{\boldsymbol{g}_k^{\mathsf{H}} \boldsymbol{K}_k^{-1} \boldsymbol{y}}{\sqrt{\mathsf{SNR}}\, \boldsymbol{g}_k^{\mathsf{H}} \boldsymbol{K}_k \boldsymbol{g}_k} - \bar{x}_k \right\|. \end{aligned}$$

The term $\frac{\boldsymbol{g}_k^{\mathsf{H}} \boldsymbol{K}_k^{-1} \boldsymbol{y}}{\sqrt{\mathsf{SNR}}\, \boldsymbol{g}_k^{\mathsf{H}} \boldsymbol{K}_k \boldsymbol{g}_k}$ is the unbiased estimator of $x_k$. In this case, the AO group decoding algorithm is equivalent to the unbiased MMSE decoding [41]. By a proper scaling, we can get the MMSE estimator from $\frac{\boldsymbol{g}_k^{\mathsf{H}} \boldsymbol{K}_k^{-1} \boldsymbol{y}}{\sqrt{\mathsf{SNR}}\, \boldsymbol{g}_k^{\mathsf{H}} \boldsymbol{K}_k \boldsymbol{g}_k}$ [41]. Although the MMSE estimator is optimal with respect to the mean squared error, it may not be optimal with respect to the symbol error probability and the unbiased MMSE may have a better performance [3].

### *C. AO-SIC Group Decoding Algorithm*

Similar to the PIC case, we can use the SIC technique to aid the AO group decoding process. The decoding order can be simply determined according to the maximum SINR criterion, which is similar to the PIC-SIC case. Suppose the ordered symbol sets are

$$\boldsymbol{x}_{\mathcal{I}_{i_0}}, \boldsymbol{x}_{\mathcal{I}_{i_1}}, \ldots, \boldsymbol{x}_{\mathcal{I}_{i_{n-1}}}. \tag{40}$$

The following is the AO-SIC group decoding algorithm:

1) Decode the first set of symbols $\boldsymbol{x}_{\mathcal{I}_{i_0}}$ using the AO group decoding algorithm;
2) Let $k=0$, $\boldsymbol{y}_0=\boldsymbol{y}$, where $\boldsymbol{y}$ is defined as in (11);
3) Remove the components of the already-detected symbol set $\boldsymbol{x}_{\mathcal{I}_{i_k}}$ from the (11),

$$\boldsymbol{y}_{k+1} \triangleq \boldsymbol{y}_k - \sqrt{\mathsf{SNR}}\boldsymbol{G}_{\mathcal{I}_k}\boldsymbol{x}_{\mathcal{I}_{i_k}} = \sqrt{\mathsf{SNR}}\sum_{j=k+1}^{N-1}\boldsymbol{G}_{\mathcal{I}_{i_j}}\boldsymbol{x}_{\mathcal{I}_{i_j}} + \boldsymbol{w}; \tag{41}$$

4) Decode $\boldsymbol{x}_{\mathcal{I}_{i_{k+1}}}$ in (41) using the AO group decoding algorithm;
5) If $k<N-1$, then set $k:=k+1$, go to Step 3; otherwise stop the algorithm.

## VI. Design Examples

In this section, we present two design examples that achieve the full diversity conditions with pair-by-pair PIC group decoding.

### A. Example 1

Consider a code for 2 transmit antennas with 3 time slots of the following form,

$$\boldsymbol{X} = \begin{bmatrix} cx_0+sx_1 & cx_2+sx_3 & 0 \\ 0 & -sx_0+cx_1 & -sx_2+cx_3 \end{bmatrix}, \tag{42}$$

where $c=\cos\theta, s=\sin\theta, \theta\in[0,2\pi)$. The symbol rate of this code is $\frac{4}{3}$.

In the following, we show that this code can be decoded with pair-by-pair PIC group decoding.

**Theorem 3.** *Let $\mathcal{A}\subset\mathbb{Z}[\boldsymbol{i}]$ be a QAM signal constellation. Let $\mathcal{I}=\{\{0,1\},\{2,3\}\}$ be a grouping scheme for the PIC group decoding algorithm. If $\tan\theta\notin\mathbb{Q}$, then code $\boldsymbol{X}$ in (42) achieves full diversity using the PIC group decoding algorithm with the grouping scheme $\mathcal{I}$.*

*Proof:* Firstly, we prove that the code given in (42) has full rank property for any $\mathcal{A}\subset\mathbb{Z}[\boldsymbol{i}]$. In order to prove this, we only need to prove that for any $x_i\in\mathbb{Z}[\boldsymbol{i}], i=0,1,2,3$, which satisfies that $x_i$ not all equal to zero, $\boldsymbol{X}$ is full rank. Since $\tan\theta\notin\mathbb{Q}$, equation $cx_0+sx_1=0$ holds for $x_0,x_1\in\Delta\mathcal{A}$ if and only if $x_0=x_1=0$. Similarly, equation $-sx_2+cx_3=0$ holds for $x_2,x_3\in\Delta\mathcal{A}$ if and only if $x_2=x_3=0$. Next, we discuss two different cases.

i). When $x_0$ and $x_1$ are not all equal to zero and $x_2$ and $x_3$ not all equal to zero, then $cx_0+sx_1\neq 0$, $-sx_2+cx_3\neq 0$. In this case, $\boldsymbol{X}$ is full rank;

ii). When $x_0$ and $x_1$ are not all equal to zero but $x_2 = x_3 = 0$, then

$$\boldsymbol{X} = \begin{bmatrix} cx_0 + sx_1 & 0 & 0 \\ 0 & -sx_0 + cx_1 & 0 \end{bmatrix}$$

is full rank; similarly, in the case when $x_2$ and $x_3$ are not all equal to zero but $x_0 = x_1 = 0$, $\boldsymbol{X}$ is full rank too.

So the code in (42) has full rank property.

Next, we prove that the code $\boldsymbol{X}$ satisfies the second condition in the main theorem. Suppose there is only one receive antenna, the equivalent channel can be written as

$$[\boldsymbol{g}_0, \boldsymbol{g}_1, \boldsymbol{g}_2, \boldsymbol{g}_3] = \begin{bmatrix} ch_0 & sh_0 & 0 & 0 \\ -sh_1 & ch_1 & ch_0 & sh_0 \\ 0 & 0 & -sh_1 & ch_1 \end{bmatrix},$$

obviously $\boldsymbol{g}_0$ and $\boldsymbol{g}_1$ can not be expressed as a linear combination of $\boldsymbol{g}_2, \boldsymbol{g}_3$, and vice versa, when $\boldsymbol{h} \neq \boldsymbol{0}$. Thus, $[\boldsymbol{g}_0, \boldsymbol{g}_1]$ and $[\boldsymbol{g}_2, \boldsymbol{g}_3]$ are linearly independent, when $\boldsymbol{h} \neq \boldsymbol{0}$. According to the main theorem, the code achieves full diversity with the PIC decoding algorithm provided that the grouping scheme is $\mathcal{I} = \{\{0,1\}, \{2,3\}\}$. ■

### B. Example 2

The following code is designed for 4 transmit antennas with 6 time slots.

$$\boldsymbol{X} = \begin{bmatrix} cx_0 + sx_1 & -cx_2^* - sx_3^* & cx_4 + sx_5 & -cx_6^* - sx_7^* & 0 & 0 \\ 0 & 0 & cx_0 + sx_1 & -cx_2^* - sx_3^* & cx_4 + sx_5 & -cx_6^* + sx_7^* \\ cx_2 + sx_3 & cx_0^* + sx_1^* & cx_6 + sx_7 & cx_4^* + sx_5^* & 0 & 0 \\ 0 & 0 & cx_2 + sx_3^* & cx_0^* + sx_1^* & cx_6 + sx_7 & cx_4^* + sx_5^* \end{bmatrix}, \quad (43)$$

where $c = \cos\theta$, $s = \sin\theta$, $\theta \in [0, 2\pi)$. It can be proved that this code satisfies the two conditions given in the main theorem if the grouping scheme is $\mathcal{I} = \{\{0,1\}, \{2,3\}, \{4,5\}, \{6,7\}\}$.

**Theorem 4.** *Let $\mathcal{A} \subset \mathbb{Z}[\boldsymbol{i}]$ be a QAM signal constellation. Let $\mathcal{I} = \{\{0,1\}, \{2,3\}, \{4,5\}, \{6,7\}\}$ be a grouping scheme for the PIC group algorithm. If $\tan\theta \notin \mathbb{Q}$, then the code $\boldsymbol{X}$ in* (43) *achieves full diversity using the PIC group decoding algorithm with the grouping scheme $\mathcal{I}$.*

*Proof:* The proof is similar to the 2-transmit-antenna case. First we prove that this code satisfies the full rank criterion. This is easy to verify just by looking into the code case by case as the previous proof.

Next we prove that the second condition in the main theorem also holds. In the case when there is only one receive antenna, the equivalent channel matrix can be written as follows,

$$\begin{aligned} \boldsymbol{G} &= [\boldsymbol{g}_0, \boldsymbol{g}_1, \boldsymbol{g}_2, \boldsymbol{g}_3, \boldsymbol{g}_4, \boldsymbol{g}_5, \boldsymbol{g}_6, \boldsymbol{g}_7] \\ &= \begin{bmatrix} ch_0 & sh_0 & ch_1 & sh_1 & 0 & 0 & 0 & 0 \\ ch_1^* & sh_1^* & -ch_0^* & -sh_0^* & 0 & 0 & 0 & 0 \\ -sh_2 & ch_2 & -sh_3 & ch_3 & ch_0 & sh_0 & ch_1 & ch_1 \\ -sh_3^* & ch_3^* & sh_2^* & -ch_2^* & ch_1^* & sh_1^* & -ch_0^* & -sh_0^* \\ 0 & 0 & 0 & 0 & -sh_2 & ch_2 & -sh_3 & ch_3 \\ 0 & 0 & 0 & 0 & -sh_3^* & ch_3^* & sh_2^* & -ch_2^* \end{bmatrix}. \end{aligned} \tag{44}$$

Let $\boldsymbol{h} \neq \boldsymbol{0}$. We can see that $[\boldsymbol{g}_0, \boldsymbol{g}_1]$ is orthogonal to $[\boldsymbol{g}_2, \boldsymbol{g}_3]$. Vector group $[\boldsymbol{g}_0, \boldsymbol{g}_1]$ is also linearly independent of $\boldsymbol{g}_4, \boldsymbol{g}_5, \boldsymbol{g}_6, \boldsymbol{g}_7$. Thus, $[\boldsymbol{g}_0, \boldsymbol{g}_1]$ can not be expressed by any linear combination of the rest column vectors in $\boldsymbol{G}$. A similar discussion can be applied to the other vector groups. Therefore, the second condition in the main theorem also holds. This completes the proof. ∎

## VII. Simulation

In this section, we present some simulation results. In all the simulations, the channel is assumed quasi-static Rayleigh flat fading. First we choose the rotation angle $\theta$ for the codes in (42) and (43) by numerically estimating the coding gains of the codes for a series of values of $\theta$. Here the coding gain $C_g$ is defined as

$$C_g = \arg\max_{\bar{C}_g} \left\{ \bar{C}_g \mid \boldsymbol{P}_{\mathsf{blockerr}}(\mathsf{SNR}) \leq \frac{1}{\bar{C}_g} \mathsf{SNR}^{-D_g} \right\}, \tag{45}$$

where $D_g$ is the diversity order. We use Monte Carlo simulations to estimate the coding gains for different $\theta$'s. As we can see from Fig. 1, the peak value of $C_g$ is reached at two points: $\theta = 0.55$ and $\theta = 1.02$. Interestingly enough, these two values of $\theta$ are very close to $\frac{1}{2}\arctan(2)$ and $\frac{\pi}{2} - \frac{1}{2}\arctan(2)$, which maximize the coding gain of the $2 \times 2$ diagonal code [41]. An intuitive explanation is that the code in (42) can be viewed as two diagonal codes stacked together and even after the interference cancellation, $\theta = \frac{1}{2}\arctan(2)$ and $\theta = \frac{\pi}{2} - \frac{1}{2}\arctan(2)$ still maximize the coding gain.

In Fig 2, we compare our new code in (42) for 2 transmit antennas with the Alamouti code for 2 transmit antennas, Golden code [2] for $2$ transmit antennas, and the QOSTBC for $4$ transmit antennas with the optimal rotation [35] and the symbol-pairwise decoding. The number of receive antennas is $3$. The constellation for our new code is $64$ QAM. The constellation for the Alamouti code and for the QOSTBC is $256$ QAM and for Golden code is $16$ QAM. Thus, in all schemes, the transmission rate is $8$ bits/sec/Hz. The PIC group decoding with group size $2$, i.e., symbol-pair-wise decoding, is used for our new code and Golden code decodings. Fig. 2 shows that our proposed coding scheme is about $1.5$ dB better than the Alamouti code and outperforms the QOSTBC before $\mathsf{SNR}$ reaches $27$ dB, while the PIC group decoding has higher complexity than the symbol-wise decoding for the Alamouti code but has similar complexity as the QOSTBC ML symbol-pair-wise decoding. For Golden code, since it is only full rank (although with non-vanishing determinant), it does not achieve full diversity when the PIC group decoding is used, which can be seen from Fig. 2.

For the $4$ by $6$ code in (43) for $4$ transmit antennas, we compare it with the QOSTBC with the optimal rotation [35] and Nguyen-Choi code [28]. The number of receive antennas is also $3$ for all these codes. Our new coding scheme uses a 64-QAM constellation and the QOSTBC uses a 256-QAM constellation so that the bit rates for both schemes are $8$ bits/sec/Hz. For Nguyen-Choi code, the constellation is 32-QAM (it is obtained by deleting the four corner points from the 6 by 6 square QAM as what is commonly used) so that the bit rate is $7.5$ bits/sec/Hz. We use the PIC and PIC-SIC group decodings for the new code, respectively, and the ML decoding for the QOSTBC, and the PIC-SIC group decoding for Nguyen-Choi code. In this case, all these decodings are symbol-pair-wise based. The simulation results show that our new code with the PIC group decoding and the PIC-SIC group decoding is $2.3$ dB and $2.8$ dB better than the QOSTBC, respectively. From Fig. 3, one can see that our new code does achieve full diversity as compared with the full diversity QOSTBC and the diversity gain of Nguyen-Choi code is smaller than that of our new code.

Note that the overlapped Alamouti codes proposed in [32], [33] with linear receivers may outperform all the other existing codes with linear receivers in the literature including perfect codes [11], [29] but do not outperform OSTBC or QOSTBC for the number of transmit antennas below $5$.

## VIII. Conclusion

In this paper, we first proposed a PIC group decoding algorithm and an AO group decoding algorithm that fill the gaps between the ML decoding algorithm and the symbol-by-symbol linear decoding algorithms namely the ZF and the MMSE decoding algorithms, respectively. We also studied their corresponding SIC-aided decoding algorithms: the PIC-SIC and the AO-SIC group decoding algorithms. We then derived a design criterion for codes to achieve full diversity when they are decoded with the PIC, AO, PIC-SIC and AO-SIC group decoding algorithms. The new derived criterion is a group independence criterion for an equivalent channel matrix and fills the gap between the loosest full rank criterion for the ML receiver and the strongest linear independence criterion of the equivalent channel matrix for linear receivers. Note that the full rank criterion is equivalent to the loosest linear independence for the column vectors of the equivalent channel matrix over a difference set of a finite signal constellation while the strongest linear independence criterion is the linear independence for the column vectors of the equivalent channel matrix over the whole complex field. The relaxed condition in the new design criterion for STBC to achieve full diversity with the PIC group decoding provides an STBC rate bridge between $n_t$ and $1$, where rate $n_t$ is the full symbol rate for the ML receiver and rate $1$ is the symbol rate upper bound for linear receivers. Thus, it provides a trade-off between decoding complexity and symbol rate when full diversity is required. We finally presented two design examples for 2 and 4 transmit antennas of rate $4/3$ that satisfy the new design criterion and thus they achieve full diversity with the PIC group decoding of group size $2$, i.e., complex-pair-wise decoding. It turns out that they may outperform the well-known Alamouti code, Golden code, and QOSTBC.

## APPENDIX A
## PROOF OF LEMMA 1

*Proof:* Writing $\boldsymbol{P}_{\mathcal{I}_k}$ defined in (15) and an arbitrary matrix $\tilde{\boldsymbol{P}}_{\mathcal{I}_k} \in \mathcal{P}_{\mathcal{I}_k}$ in the following forms,

$$\boldsymbol{P}_{\mathcal{I}_k} = \left[\boldsymbol{p}_0^{\mathsf{T}}, \boldsymbol{p}_1^{\mathsf{T}}, \ldots, \boldsymbol{p}_{m-1}^{\mathsf{T}}\right]^{\mathsf{T}},$$

$$\tilde{\boldsymbol{P}}_{\mathcal{I}_k} = \left[\tilde{\boldsymbol{p}}_0^{\mathsf{T}}, \tilde{\boldsymbol{p}}_1^{\mathsf{T}}, \ldots, \tilde{\boldsymbol{p}}_{m-1}^{\mathsf{T}}\right]^{\mathsf{T}},$$

according to the definition of $\mathcal{P}_{\mathcal{I}_k}$ in (18), we must have that $\boldsymbol{p}_i^*, \tilde{\boldsymbol{p}}_i^* \in \mathbb{V}_{\mathcal{I}_k}^{\perp}$, $i = 0, 1, \ldots, m-1$. Note that $\operatorname{rank}\left(\boldsymbol{P}_{\mathcal{I}_k}^*\right) = \operatorname{rank}\left(\boldsymbol{P}_{\mathcal{I}_k}\right) = \dim\left(\mathbb{V}_{\mathcal{I}_k}^{\perp}\right)$, which implies that all the vectors in $\mathbb{V}_{\mathcal{I}_k}^{\perp}$ can be expressed as linear combinations of $\boldsymbol{p}_i^*, i = 0, 1, \ldots, m-1$. So there must exist $f_{i,j}^*, 0 \leq i, j < m$,

such that

$$\tilde{\boldsymbol{p}}_i^* = \sum_{i=0}^{m-1} f_{i,j}^* \boldsymbol{p}_j^*,$$

or in the matrix form we have $\tilde{\boldsymbol{P}}_{\mathcal{I}_k} = \boldsymbol{F}\boldsymbol{P}_{\mathcal{I}_k}$, where the $(i,j)$-th entry of $\boldsymbol{F}$ is $f_{i,j}$. So $\tilde{\boldsymbol{P}}_{\mathcal{I}_k}$ can be viewed as a concatenation of the linear filters $\boldsymbol{P}_{\mathcal{I}_k}$ and $\boldsymbol{F}$. Substituting the above equation into (19), we get

$$\tilde{\boldsymbol{z}}_{\mathcal{I}_k} = \boldsymbol{F}\left(\boldsymbol{P}_{\mathcal{I}_k}\boldsymbol{G}_{\mathcal{I}_k}\boldsymbol{x}_{\mathcal{I}_k} + \boldsymbol{P}_{\mathcal{I}_k}\boldsymbol{w}\right) = \boldsymbol{F}\boldsymbol{z}_{\mathcal{I}_k}, \tag{46}$$

where $\boldsymbol{z}_{\mathcal{I}_k} = \boldsymbol{P}_{\mathcal{I}_k}\boldsymbol{G}_{\mathcal{I}_k}\boldsymbol{x}_{\mathcal{I}_k} + \boldsymbol{P}_{\mathcal{I}_k}\boldsymbol{w}$. For an $\mathsf{SNR}$, the optimal decoding of $\boldsymbol{x}_{\mathcal{I}_k}$ from $\boldsymbol{z}_{\mathcal{I}_k}$ is as follows,

$$\hat{\boldsymbol{x}}_{\mathcal{I}_k} = \arg\min_{\bar{\boldsymbol{x}}_{\mathcal{I}_k} \in \mathcal{A}^{n_k}} P\left(\boldsymbol{z}_{\mathcal{I}_k} \middle| \bar{\boldsymbol{x}}_{\mathcal{I}_k}\right),$$

and the optimal decoding of $\boldsymbol{x}_{\mathcal{I}_k}$ from $\tilde{\boldsymbol{z}}_{\mathcal{I}_k}$ is as follows,

$$\hat{\boldsymbol{x}}_{\mathcal{I}_k} = \arg\min_{\bar{\boldsymbol{x}}_{\mathcal{I}_k} \in \mathcal{A}^{n_k}} P\left(\boldsymbol{F}\boldsymbol{z}_{\mathcal{I}_k} \middle| \bar{\boldsymbol{x}}_{\mathcal{I}_k}\right). \tag{47}$$

Notice that any filtering may not help an ML decision. Therefore, for an $\mathsf{SNR}$, we have

$$P_{\mathsf{err}}(\boldsymbol{P}_{\mathcal{I}_k}, \mathsf{SNR}) \leq P_{\mathsf{err}}(\tilde{\boldsymbol{P}}_{\mathcal{I}_k}, \mathsf{SNR}),$$

which completes the proof. ■

## Appendix B
## Proof of Lemma 2

*Proof:* Since $\boldsymbol{P}$ is a projection matrix, $\boldsymbol{P}$ can be decomposed as

$$\boldsymbol{P} = \boldsymbol{U}^{\mathsf{H}}\boldsymbol{D}\boldsymbol{U}, \tag{48}$$

where $\boldsymbol{U} \in \mathbb{C}^{m\times m}$ is an unitary matrix and

$$\boldsymbol{D} = \begin{bmatrix} \boldsymbol{I}_{r\times r} & \boldsymbol{0}_{r\times m-r} \\ \boldsymbol{0}_{m-r\times r} & \boldsymbol{0}_{m-r\times m-r} \end{bmatrix}, \quad r = \mathrm{rank}(\boldsymbol{P}). \tag{49}$$

By multiplying both sides of (20) by $\boldsymbol{U}$ to the left, we have

$$\boldsymbol{U}\boldsymbol{y} = \boldsymbol{U}\boldsymbol{G}\boldsymbol{x} + \boldsymbol{D}\boldsymbol{U}\boldsymbol{w}. \tag{50}$$

Since the column vectors of $\boldsymbol{G}$ belong to $\mathbb{V}$, $\boldsymbol{G} = \boldsymbol{P}\boldsymbol{G}$, (50) can be written as

$$\boldsymbol{U}\boldsymbol{y} = \boldsymbol{U}\boldsymbol{P}\boldsymbol{G}\boldsymbol{x} + \boldsymbol{D}\boldsymbol{U}\boldsymbol{w} = \boldsymbol{D}\boldsymbol{U}\boldsymbol{G}\boldsymbol{x} + \boldsymbol{D}\boldsymbol{U}\boldsymbol{w} \tag{51}$$

Note that the effect of multiplying $\boldsymbol{D}$ to the left of a vector is picking up the first $r$ entries and setting the rest $n-r$ entries to zero. Hence from (51), we can see that only the first $r$ entries of $\boldsymbol{Uy}$ matter and all other entries are zeros. We also have that the first $r$ entries of $\boldsymbol{DUw}$ are i.i.d. Gaussian noise since $\boldsymbol{U}$ is unitary, the rest $n-r$ entries are all zeros. Use $[\boldsymbol{v}]_r$ to denote the vector that contains the first $r$ entries of $\boldsymbol{v}\in\mathbb{C}^m$. Then, (51) is equivalent to

$$[\boldsymbol{Uy}]_r = [\boldsymbol{DUGx}]_r + [\boldsymbol{DUw}]_r. \tag{52}$$

Since $[\boldsymbol{DUw}]_r$ is a white Gaussian noise, the ML decision is the same as the minimum distance decision for (52), i.e.,

$$\begin{aligned}\hat{\boldsymbol{x}} &= \arg\min_{\bar{\boldsymbol{x}}\in\mathcal{A}} \left\|[\boldsymbol{Uy}]_r - [\boldsymbol{DUG}\bar{\boldsymbol{x}}]_r\right\| \\ &= \arg\min_{\bar{\boldsymbol{x}}\in\mathcal{A}} \left\|\boldsymbol{Uy} - \boldsymbol{DUG}\bar{\boldsymbol{x}}\right\|,\end{aligned} \tag{53}$$

where the second equality holds because the last $n-r$ entries have no effect on the distance. Noting that $\boldsymbol{U}$ is an unitary matrix and $\boldsymbol{G}=\boldsymbol{PG}$, the above detection is equivalent to

$$\hat{\boldsymbol{x}} = \arg\min_{\bar{\boldsymbol{x}}\in\mathcal{A}^n} \left\|\boldsymbol{U}^{\mathsf{H}}\boldsymbol{Uy} - \boldsymbol{U}^{\mathsf{H}}\boldsymbol{DUG}\bar{\boldsymbol{x}}\right\| = \arg\min_{\bar{\boldsymbol{x}}\in\mathcal{A}^n} \left\|\boldsymbol{y} - \boldsymbol{G}\bar{\boldsymbol{x}}\right\|. \tag{54}$$

Thus, we conclude that the minimum distance decision in this case is equivalent to the maximum likelihood decision. ■

## APPENDIX C
## PROOF OF LEMMA 3

*Proof:* For a given $\boldsymbol{h}=[h_0,h_1,\ldots,h_{l-1}]^{\mathsf{T}}$, and two symbol vectors $\boldsymbol{x},\widetilde{\boldsymbol{x}}\in\mathcal{A}^n$ with $\boldsymbol{x}\neq\widetilde{\boldsymbol{x}}$, the pairwise error probability $P_{\boldsymbol{h}}(\boldsymbol{x}\to\widetilde{\boldsymbol{x}})$ with ML receiver is as follows,

$$\begin{aligned}P_{\boldsymbol{h}}(\boldsymbol{x}\to\widetilde{\boldsymbol{x}}) =& Q\left(\sqrt{\mathsf{SNR}}\left\|\boldsymbol{G}(\boldsymbol{h})\Delta\boldsymbol{x}\right\|\right) \\ \leq& Q\left(c\cdot\sqrt{\mathsf{SNR}}\left(\sum_{i=0}^{l-1}|h_i|^2\right)^{\frac{1}{2}}\left\|\Delta\boldsymbol{x}\right\|\right) \\ \leq& \frac{1}{2}\exp\left(-\frac{c^2\cdot\mathsf{SNR}}{2}\sum_{i=0}^{l-1}|h_i|^2\left\|\Delta\boldsymbol{x}\right\|^2\right),\end{aligned} \tag{55}$$

where the last inequality is obtained by applying the well-known upper-bound for the $Q$-function,

$$Q(x)\leq\frac{1}{2}\exp\left(-\frac{x^2}{2}\right).$$

By taking expectation over $\boldsymbol{h}$ at both sides of (55), we get

$$P\left(\boldsymbol{x}\to\widetilde{\boldsymbol{x}}\right)=\mathcal{E}_{\boldsymbol{h}}\left\{P_{\boldsymbol{h}}(\boldsymbol{x}\to\widetilde{\boldsymbol{x}})\right\}$$
$$\leq\mathcal{E}_{\boldsymbol{h}}\left\{\exp\left(-\frac{c^2\cdot\mathsf{SNR}}{2}\sum_{i=0}^{l-1}\left|h_i\right|^2\left\|\Delta\boldsymbol{x}\right\|^2\right)\right\}.$$

To evaluate the above expectation, we use

$$\mathcal{E}_h\left(\exp(-a\left|h\right|^2\right)=\frac{1}{1+a},\ h\sim\mathcal{CN}\left(0,1\right),\ a>0,$$

and note that the expectation can be taken separately to each $h_i$, which leads to the following result,

$$P(\boldsymbol{x}\to\tilde{\boldsymbol{x}})\leq\frac{1}{2}\left(\frac{2}{2+c^2\mathsf{SNR}\left\|\Delta\boldsymbol{x}\right\|^2}\right)^l$$
$$<\frac{2^{l-1}}{c^{2l}\left\|\Delta\boldsymbol{x}\right\|^{2l}}\mathsf{SNR}^{-l}.$$

Since $\Delta\boldsymbol{x}\in\Delta\mathcal{A}^n$ and $\Delta\mathcal{A}^n$ is a finite set, there exists a $\Delta\boldsymbol{x}_0$ such that

$$d_{\min}=\left\|\Delta\boldsymbol{x}_0\right\|=\min\left\{\left\|\Delta\boldsymbol{x}\right\|,\mathbf{0}\neq\Delta\boldsymbol{x}\in\Delta\mathcal{A}^n\right\}.$$

Hence for any $\boldsymbol{x},\widetilde{\boldsymbol{x}}\in\mathcal{A}^n$ with $\boldsymbol{x}\neq\widetilde{\boldsymbol{x}}$, we always have

$$P(\boldsymbol{x}\to\tilde{\boldsymbol{x}})<\frac{2^{l-1}}{\left(c\cdot d_{\min}\right)^{2l}}\mathsf{SNR}^{-l}.$$

The symbol error probability $P_{\mathsf{SER}}(\mathsf{SNR})$ is upper-bounded by

$$P_{\mathsf{SER}}(\mathsf{SNR})<\frac{2^{l-1}\left(\left|\mathcal{A}\right|^n-1\right)}{\left(c\cdot d_{\min}\right)^{2l}}\mathsf{SNR}^{-l},$$

i.e., the system achieves the diversity order $l$. ■

## Appendix D
## Proof of Lemma 4

*Proof:* For a given $\mathbf{0}\neq\boldsymbol{h}\in\mathbb{C}^l$ and $\mathbf{0}\neq\Delta\boldsymbol{x}\in\Delta\mathcal{A}^{n_k}$, since the column vectors of $\boldsymbol{G}(\boldsymbol{h})$ are linearly independent over $\Delta\mathcal{A}$, $\boldsymbol{G}(\boldsymbol{h})\Delta\boldsymbol{x}\neq\mathbf{0}$, or

$$\left\|\boldsymbol{G}(\boldsymbol{h})\Delta\boldsymbol{x}\right\|>0. \tag{56}$$

Now let us consider a fixed $\Delta\boldsymbol{x}\in\Delta\mathcal{A}^n$ and restrict the parameter $\boldsymbol{h}$ to the unit sphere, i.e., $\left\|\boldsymbol{h}\right\|=1$. Since the unit sphere is a compact set, from (56), for this $\Delta\boldsymbol{x}$ there must exist a constant $c_{\Delta\boldsymbol{x}}>0$ such that

$$\left\|\boldsymbol{G}(\boldsymbol{h})\Delta\boldsymbol{x}\right\|\geq c_{\Delta\boldsymbol{x}}. \tag{57}$$

For $\mathbf{0} \neq \boldsymbol{h} \in \mathbb{C}^l$, we always have

$$\left\| \frac{1}{\|\boldsymbol{h}\|} \boldsymbol{G}(\boldsymbol{h}) \Delta \boldsymbol{x} \right\| = \left\| \boldsymbol{G}\left( \frac{\boldsymbol{h}}{\|\boldsymbol{h}\|} \right) \Delta \boldsymbol{x} \right\| \geq c_{\Delta \boldsymbol{x}}, \tag{58}$$

or

$$\|\boldsymbol{G}(\boldsymbol{h}) \Delta \boldsymbol{x}\| \geq c_{\Delta \boldsymbol{x}} \|\boldsymbol{h}\| = c_{\Delta \boldsymbol{x}} \left( \sum_{i=0}^{l-1} |h_i|^2 \right)^{\frac{1}{2}}. \tag{59}$$

Since $\Delta \mathcal{A}^n$ is a finite set, we can define $c_{\min}$ and $d_{\max}$ so that

$$0 < c_{\min} = \min \left\{ c_{\Delta \boldsymbol{x}}, \mathbf{0} \neq \Delta \boldsymbol{x} \in \Delta \mathcal{A}^n \right\}, \tag{60}$$

$$0 < d_{\max} = \max \left\{ \|\Delta \boldsymbol{x}\|, \Delta \boldsymbol{x} \in \Delta \mathcal{A}^n \right\}. \tag{61}$$

Then

$$\begin{aligned} \|\boldsymbol{G}(\boldsymbol{h}) \Delta \boldsymbol{x}\| &\geq \frac{c_{\min}}{d_{\max}} \left( \sum_{i=0}^{l-1} |h_i|^2 \right)^{\frac{1}{2}} \|\Delta \boldsymbol{x}\| \\ &= c \left( \sum_{i=0}^{l-1} |h_i|^2 \right)^{\frac{1}{2}} \|\Delta \boldsymbol{x}\|, \ \forall \Delta \boldsymbol{x} \in \Delta \mathcal{A}^n, \boldsymbol{h} \in \mathbb{C}^l, \end{aligned} \tag{62}$$

where $c \triangleq \frac{c_{\min}}{d_{\max}}$. This completes the proof. ■

## APPENDIX E
## PROOF OF THEOREM 2

*Proof:* Let $\boldsymbol{H} = (h_{i,j}) \in \mathbb{C}^{n_t \times n_r}$ be the channel matrix as in (1) and $\boldsymbol{h} = \mathrm{vec}(\boldsymbol{H})$. Suppose $\mathcal{X}$ is an STBC that satisfies the full rank criterion, i.e., any matrix $\mathbf{0} \neq \Delta \boldsymbol{X} \in \Delta \mathcal{X}$ is a full rank matrix. Write $\Delta \boldsymbol{X} \Delta \boldsymbol{X}^{\mathsf{H}}$ into the following decomposition

$$\Delta \boldsymbol{X} \Delta \boldsymbol{X}^{\mathsf{H}} = \boldsymbol{U} \boldsymbol{D} \boldsymbol{U}^{\mathsf{H}}, \tag{63}$$

where $\boldsymbol{U} \in \mathbb{C}^{n_t \times n_t}$ is an unitary matrix and $\boldsymbol{D} = \mathrm{diag}(\lambda_0, \lambda_1, \ldots, \lambda_{n_t - 1})$. Since $\Delta \boldsymbol{X}$ is a full rank matrix, $\lambda_{\min}(\Delta \boldsymbol{X}) \triangleq \min \{\lambda_0, \lambda_1, \ldots, \lambda_{n_t-1}\} > 0$. Note that $\Delta \boldsymbol{X}$ is a finite set, we can define $\lambda_{\min}$ such that

$$\lambda_{\min} = \min \left\{ \lambda_{\min}(\Delta \boldsymbol{X}), \mathbf{0} \neq \Delta \boldsymbol{X} \in \Delta \mathcal{X} \right\} > 0. \tag{64}$$

Hence we have

$$\begin{aligned}\|\boldsymbol{H}\Delta\boldsymbol{X}\|^2 &= \operatorname{tr}\left(\boldsymbol{H}\Delta\boldsymbol{X}\Delta\boldsymbol{X}^{\mathsf{H}}\boldsymbol{H}^{\mathsf{H}}\right)\\ &= \operatorname{tr}\left(\boldsymbol{H}\boldsymbol{U}\boldsymbol{D}\left(\boldsymbol{H}\boldsymbol{U}\right)^{\mathsf{H}}\right)\\ &\geq \operatorname{tr}\left(\lambda_{\min}\boldsymbol{H}\boldsymbol{U}\left(\boldsymbol{H}\boldsymbol{U}\right)^{\mathsf{H}}\right)\\ &= \lambda_{\min}\sum_{i=0}^{n_r-1}\sum_{j=0}^{n_t-1}\left|h_{ij}\right|^2, \forall\boldsymbol{H}\in\mathbb{C}^{n_r\times n_t}.\end{aligned} \tag{65}$$

As (7) mentioned in Section II, $\|\boldsymbol{H}\Delta\boldsymbol{X}\|^2$ can also be written as

$$\|\boldsymbol{H}\Delta\boldsymbol{X}\|^2 = \|\boldsymbol{G}(\boldsymbol{h})\Delta\boldsymbol{x}\|^2. \tag{66}$$

where $\Delta\boldsymbol{x}\in\Delta\mathcal{A}^n$. By (65) and (66), we can see that for $\boldsymbol{h}\neq\boldsymbol{0}$, $\boldsymbol{G}(\boldsymbol{h})\Delta\boldsymbol{x}=\boldsymbol{0}$ if and only if $\Delta\boldsymbol{x}=\boldsymbol{0}$, i.e., the column vectors of $\boldsymbol{G}(\boldsymbol{h})$ are linearly independent over $\Delta\mathcal{A}$.

We now prove the necessity. Since $\mathcal{X}$ is a linear dispersion code, the scaling invariance (31) is satisfied. If the column vectors of $\boldsymbol{G}(\boldsymbol{h})$ are linearly independent over $\Delta\mathcal{A}$, then according to Lemma 4, there exists a constant $c>0$ such that

$$\begin{aligned}\|\boldsymbol{H}\Delta\boldsymbol{X}\|^2 &= \operatorname{tr}\left(\boldsymbol{H}\Delta\boldsymbol{X}\Delta\boldsymbol{X}^{\mathsf{H}}\boldsymbol{H}^{\mathsf{H}}\right)\\ &\geq c\,\|\Delta\boldsymbol{x}\|^2\sum_{i=0}^{n_r-1}\sum_{j=0}^{n_t-1}\left|h_{ij}\right|^2,\ \forall\boldsymbol{H}\in\mathbb{C}^{n_r\times n_t}.\end{aligned} \tag{67}$$

Next we prove that the above inequality implies that the eigenvalues of $\Delta\boldsymbol{X}\Delta\boldsymbol{X}^{\mathsf{H}}$ are all greater than zero for $\Delta\boldsymbol{X}\neq\boldsymbol{0}$. The uniqueness from the decodablity of the STBC $\mathcal{X}$ tells us that $\Delta\boldsymbol{X}\neq\boldsymbol{0}$ implies $\Delta\boldsymbol{x}\neq\boldsymbol{0}$. Consider the decomposition (63) for $\Delta\boldsymbol{X}$. If there is an eigenvalue $\lambda_k=0$, then we can find an $\boldsymbol{H}\in\mathbb{C}^{n_r\times n_t}$ such that

$$\boldsymbol{H}\boldsymbol{U} = \left[\boldsymbol{0},\boldsymbol{0},\ldots,\boldsymbol{v},\ldots,\boldsymbol{0}\right], \tag{68}$$

where the $k$-th column vector $\boldsymbol{v}\in\mathbb{C}^{n_r}$ can be arbitrary non-zero vector. The existence of such $\boldsymbol{H}\neq\boldsymbol{0}$ is ensured since $\boldsymbol{U}$ is invertible. For the $\boldsymbol{H}$ that satisfies (68),

$$\operatorname{tr}\left(\boldsymbol{H}\Delta\boldsymbol{X}\Delta\boldsymbol{X}^{\mathsf{H}}\boldsymbol{H}^{\mathsf{H}}\right) = \operatorname{tr}\left(\boldsymbol{H}\boldsymbol{U}\boldsymbol{D}\left(\boldsymbol{H}\boldsymbol{U}\right)^{\mathsf{H}}\right) = 0, \tag{69}$$

which contradicts with the inequality in (67). So we have proved that all the eigenvalues of $\Delta\boldsymbol{X}\Delta\boldsymbol{X}^{\mathsf{H}}$ must satisfy $\lambda_i>0$, i.e., $\Delta\boldsymbol{X}$ is a full rank matrix. ∎

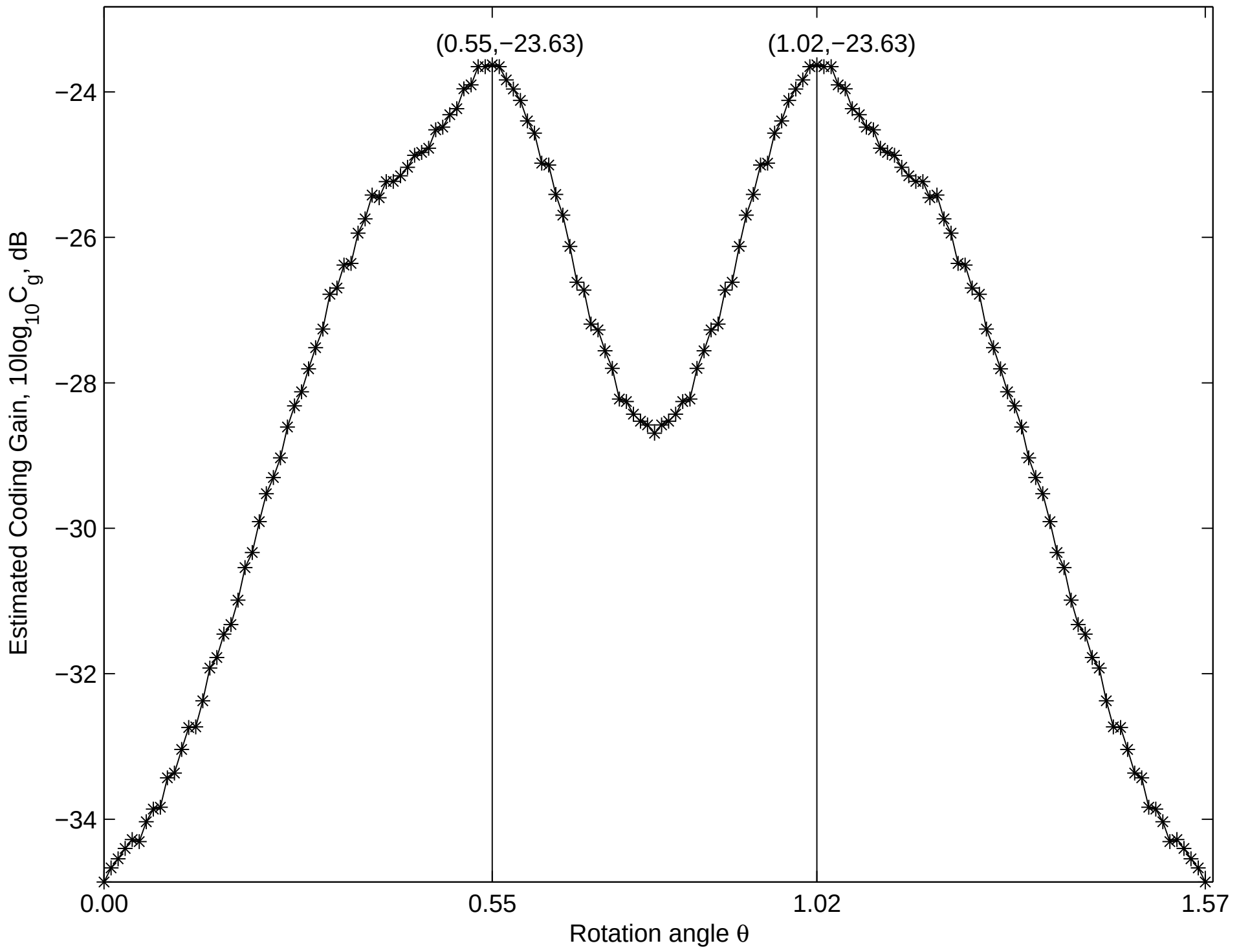

Fig. 1. Coding gain estimation for $\theta \in [0, \frac{\pi}{2})$

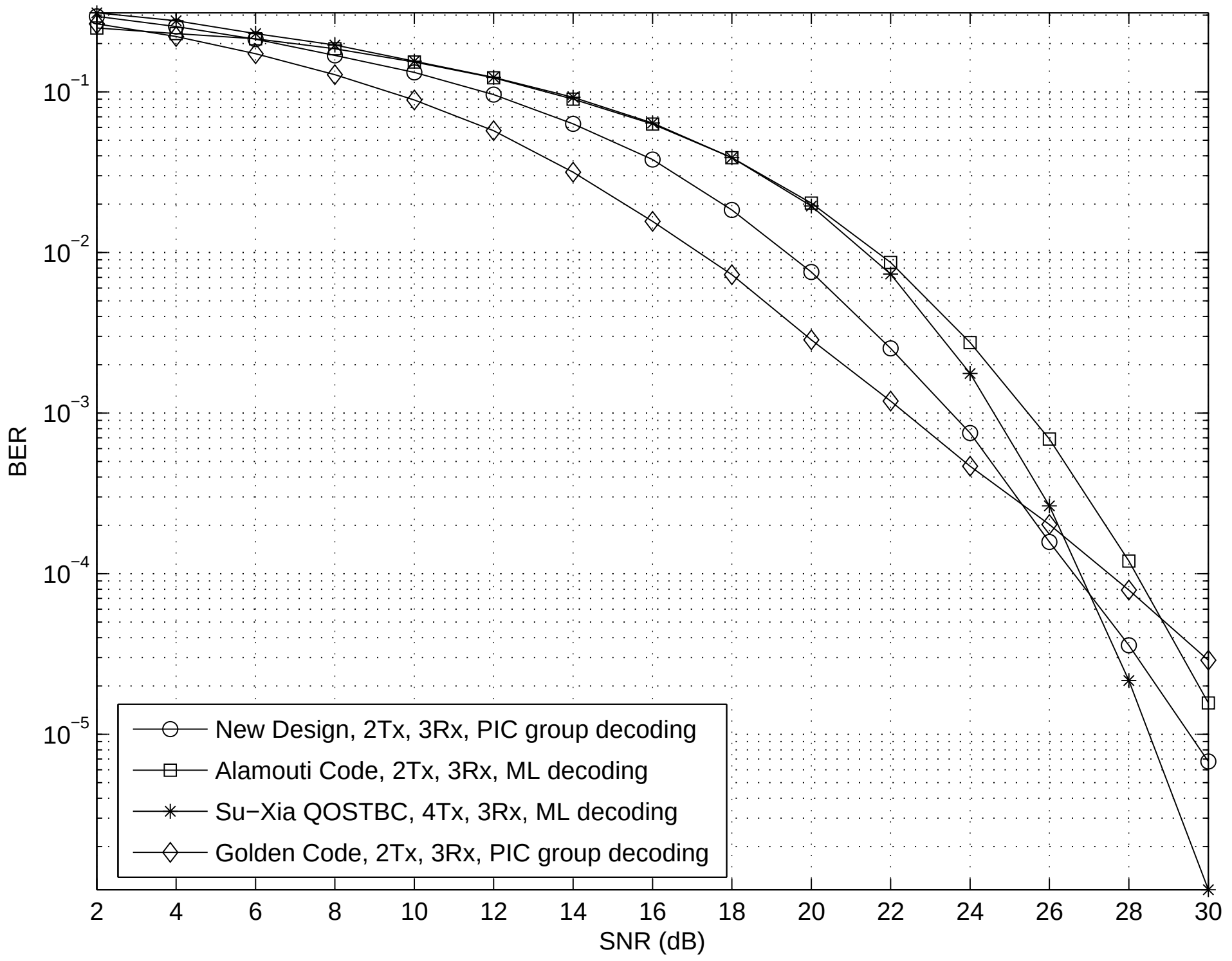

Fig. 2. Performance comparison of several coding scheme and their bandwidth efficiencies are all 8 bits/sec/Hz

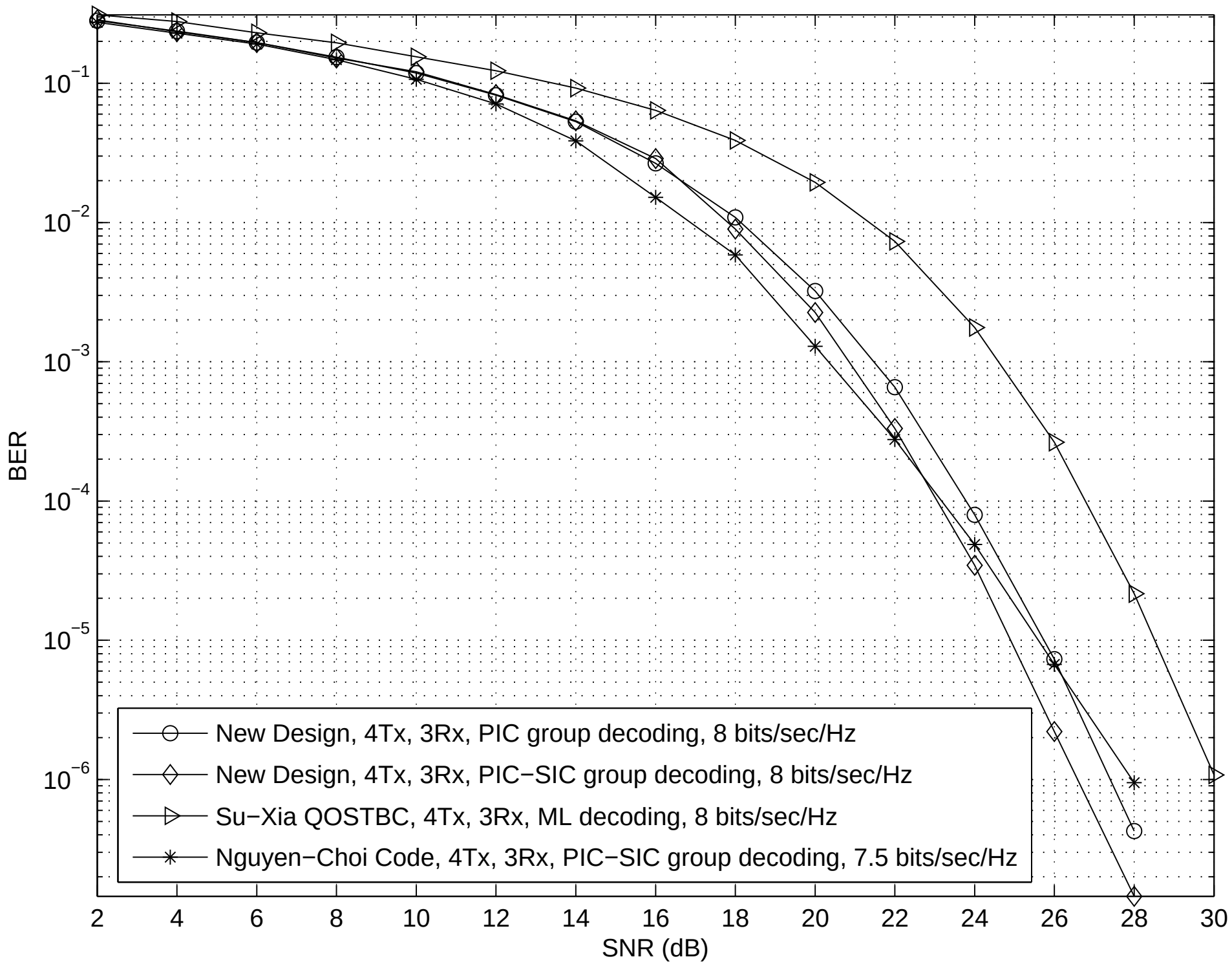

Fig. 3. Performance comparison of several coding schemes for 4 transmit and 3 receive antennas